\documentclass[a4paper,11pt]{article}
\pdfoutput=1
\usepackage{jheppub}

\usepackage{graphicx}
\usepackage{placeins}
\usepackage{url}
\usepackage{multirow}
\usepackage{amssymb}
\usepackage{amsmath}
\usepackage{xspace}
\usepackage{booktabs}
\usepackage[symbol]{footmisc}

\begin{document}

\title{Designing Observables for Measurements \\ with Deep Learning}

\affiliation[a]{Department of Physics and Astronomy, University of California, Riverside, CA 92521, USA}
\affiliation[b]{Physics Division, Lawrence Berkeley National Laboratory, Berkeley, CA 94720, USA}
\affiliation[c]{Berkeley Institute for Data Science, University of California, Berkeley, CA 94720, USA}

\author[a]{Owen Long}
\author[b,c]{and Benjamin Nachman}

\abstract{
Many analyses in particle and nuclear physics use simulations to infer fundamental, effective, or phenomenological parameters of the underlying physics models.  When the inference is performed with unfolded cross sections, the observables are designed using physics intuition and heuristics.  We propose to design targeted observables with machine learning.  Unfolded, differential cross sections in a neural network output contain the most information about parameters of interest and can be well-measured by construction.  The networks are trained using a custom loss function that rewards outputs that are sensitive to the parameter(s) of interest while simultaneously penalizing outputs that are different between particle-level and detector-level (to minimize detector distortions).  We demonstrate this idea in simulation using two physics models for inclusive measurements in deep inelastic scattering.  We find that the new approach is more sensitive than classical observables at distinguishing the two models and also has a reduced unfolding uncertainty due to the reduced detector distortions.
}

\maketitle

\section{Introduction}
\label{sec:intro}

Simulations are widely used for parameter estimation in particle and nuclear physics.  A typical analysis will follow one of two paths: forward-folding or unfolding.  In the forward-folding pipeline, the target physics model must be specified at the time of inference.  We focus on unfolding, where detector distortions are corrected in a first step (unfolding) and then the resulting cross section can be analyzed in the context of many models by any end users.  In the unfolding pipeline, the first step is to identify observables sensitive to a given parameter(s).  These are typically identified using physical reasoning.  Then, the differential cross sections of these observables are measured, which includes unfolding with uncertainty quantification.  Finally, the measured cross sections are fit to simulation templates with different values of the target parameters.  This approach has been deployed to measure fundamental parameters like the top quark mass~\cite{CMS:2022kqg} and the strong coupling constant $\alpha_s(m_Z)$~\cite{H1:2017bml, Collaboration:2023xha} as well as parton distribution functions~\cite{H1:2015ubc,Harland-Lang:2016yfn,Hou:2019efy,NNPDF:2021njg} and effective or phenomenological parameters in parton shower Monte Carlo programs~\cite{Campbell:2022qmc}.

A key drawback of the standard pipeline is that the observables are constructed manually.  There is no guarantee that the observables are maximally sensitive to the target parameters.  Additionally, the observables are usually chosen based on particle-level information alone and so detector distortions may not be small.  Such distortions can reduce the sensitivity to the target parameter once they are corrected for by unfolding.  In some cases, the particle-level observable must be chosen manually because it must be calculable precisely in perturbation theory; this is usually not the case when Monte Carlo simulations are used for the entire statistical analysis.  There have been proposals to optimize the detector-level observable for a given particle-level observable~\cite{Arratia:2022wny} since they do not need to be the same.  Alternatively, one could measure the full phase space and project out the desired observable after the fact~\cite{Datta:2018mwd,bunse2018unification,Ruhe2019MiningFS,Andreassen:2019cjw,Bellagente:2019uyp,1800956,Vandegar:2020yvw,Andreassen:2021zzk,Howard:2021pos,Backes:2022vmn,Chan:2023tbf,Shmakov:2023kjj,Alghamdi:2023emm,H1:2021wkz,H1prelim-22-031,H1:2023fzk,H1prelim-21-031,LHCb:2022rky,Komiske:2022vxg,Song:2023sxb} (see Ref.~\cite{Arratia:2021otl} for an overview).  

We propose to use machine learning for designing observables that are maximally sensitive to a given parameter(s)
or model discrimination 
while also being minimally sensitive to detector distortions.  Simultaneous optimization ensures that we only use regions of phase space that are measurable.  A tailored loss function is used to train neural networks.  We envision that this approach could be used for any case where simulations are used for parameter estimation.  For concreteness, we demonstrate the new technique to the case of differentiating two parton shower Monte Carlo models of deep inelastic scattering.  While neither model is expected to match data exactly, the availability of many events with corresponding detailed simulations makes this a useful benchmark problem.  We do not focus on the unfolding or parameter estimation steps themselves, but there are many proposals for doing unfolding~\cite{Arratia:2021otl,Datta:2018mwd,bunse2018unification,Ruhe2019MiningFS,Andreassen:2019cjw,Bellagente:2019uyp,1800956,Vandegar:2020yvw,Andreassen:2021zzk,Howard:2021pos,Backes:2022vmn,Chan:2023tbf,Shmakov:2023kjj,Alghamdi:2023emm,H1:2021wkz,H1prelim-22-031,H1:2023fzk,H1prelim-21-031,LHCb:2022rky,Komiske:2022vxg,Song:2023sxb} and parameter estimation~\cite{Brehmer:2018kdj,Brehmer:2018eca,Brehmer:2018hga,DeCastro:2018psv,Elwood:2018qsr,Andreassen:2019nnm,Brehmer:2019xox,Wunsch:2020iuh,Ghosh:2021roe,GomezAmbrosio:2022mpm,Simpson:2022suz} with machine learning.  Instead, our focus is on the construction of observables that are engineered to be sensitive to target parameters 
or to distinguishing models 
while also insensitive to detector effects.  The latter quality ensures that uncertainties arising from the dependence on unfolding `priors' is small.  This is explicitly illustrated using a standard, binned unfolding method in the deep inelastic scattering demonstration.

This paper is organized as follows.  Section~\ref{sec:methods} introduces our approach to observable construction.  The datasets used for demonstrating the new method are introduced in Sec.~\ref{sec:data}.  Results with these datasets are presented in Sec.~\ref{sec:results}.  The paper ends with conclusions and outlook in Sec.~\ref{sec:conclusion}.


\section{Methodology}
\label{sec:methods}

We begin by constructing new observables that are simultaneously sensitive to a parameter while also being minimally sensitive to detector effects.  This is accomplished by training neural networks $f$ with the following loss function:
\begin{align}
\label{eq: idea}
    L[f] = L_\text{classic}[f(z),\mu]+\lambda\, L_\text{new}[f(x),f(z)]\,,
\end{align}
where $\lambda>0$ is a hyperparameter that controls how much we regularize the network.  We have pairs of inputs $(Z,X)$ where $Z$ represents the particle-level inputs and $X$ represents the detector-level inputs.
Capital letters represent random variables while lower-case letter represent realizations of the corresponding random variables.  We consider the case $X$ and $Z$ have the same structure, i.e. they are both sets of 4-vectors (so it makes sense to compute $f(z)$ and $f(x)$).  This is the standard case where $X$ is a set of energy-flow objects that are meant to correspond to the 4-vectors of particles before being distorted by the detector.  Furthermore, we fix the same definition of the observable at particle and detector level.  

The first term in Eq.~\ref{eq: idea}, $L_\text{classic}$, governs the sensitivity of the observable $f$ to the target parameter $\mu$ at particle level.  For regression tasks, $\mu$ will be a real number, representing e.g. a (dimensionless) mass or coupling.  For two-sample tests, $\mu\in\{0,1\}$, where $0$ represents the null hypothesis and $1$ represents the alternative hypothesis.  A classification setup may also be useful for a regression task, by using two samples at different values of the target parameter.  The second term in Eq.~\ref{eq: idea}, $L_\text{new}$, governs how sensitive the new observable is to detector effects.  It has the property that it is small when $f(x)$ and $f(z)$ are the same and large otherwise.  When $\lambda\rightarrow\infty$, the observable is completely insensitive to detector effects.  This means that any uncertainty associated with removing such effects (including the dependence on unfolding `priors') is eliminated.  The best value of $\lambda$ will be problem specific and should ideally be chosen based on one or more downstream tasks with the unfolded data.

We introduce the method with a toy model for continuous parameter estimation (Sec.~\ref{sec:toy}), which demonstrates the essential ideas in a simplified context.
This is followed by a more complete binary classification example using simulated deep inelastic scattering events from the H1 experiment at HERA (Sec.~\ref{sec:binarycase}), where the goal is to be maximally sensitive to distinguishing two datasets.

\subsection{ Toy example for continuous parameter estimation}
\label{sec:toy}

The training samples are generated with a uniform distribution for the parameter of interest $\mu$, so each event is specified by $(\mu_i,z_i,x_i)$.
Then, we parameterize the observable $f$ as a neural network and optimize the following loss function:
\begin{align}
\label{eq:loss-regression}
    L[f] = \sum_{i} (f(z_i) - \mu_i)^2
    + \lambda\sum_{i} (f(x_i)-f(z_i))^2\,,
\end{align}
where the form of both terms is the usual mean squared error loss used in regression tasks.  The first term trains the regression to predict the parameter of interest $\mu$ while the second term trains the network to make the predictions given detector level features $x$ and particle level features $z$ similar.
We use the prediction based on particle-level features $z_i$ in the first term in the loss function.  Results for the alternative choice, using the detector-level features $x_i$ are shown in Appendix~\ref{alt-loss}.

The loss function in Eq.~\ref{eq:loss-regression} is similar to the setting of decorrelation, where a classifier is trained to be independent from a given feature~\cite{Blance:2019ibf,Englert:2018cfo,Louppe:2016ylz,Dolen:2016kst,Moult:2017okx,Stevens:2013dya,Shimmin:2017mfk,Bradshaw:2019ipy,ATL-PHYS-PUB-2018-014,DiscoFever,Wunsch:2019qbo,Rogozhnikov:2014zea,10.1088/2632-2153/ab9023,Kasieczka:2020pil,Kitouni:2020xgb,Estrade:2019gzk,Aguilar-Saavedra:2017rzt,aguilarsaavedra2020mass,Aguilar-Saavedra:2023pde}.  One could apply decorrelation techniques in this case to ensure the classifier is not able to distinguish between features at detector level and at particle level.  However, this will only ensure that the probability density for $f$ is the same for particle level and detector level.  To be well-measured, we need more than statistical similarity between distributions - we need them to be similar event by event.  The final term in Eq.~\ref{eq:loss-regression} is designed for exactly this purpose.

All deep neural networks are implemented in \textsc{Keras}~\cite{chollet2015keras}/\textsc{TensorFlow}~\cite{tensorflow} and optimized using \textsc{Adam}~\cite{adam}.  The network models use two hidden layers with 50 nodes per layer and Rectified Linear Unit (ReLU) activation functions for intermediate layers and a linear activation function for the last layer.

\begin{figure}[h]
   \begin{center}
\includegraphics[width=0.95\linewidth]{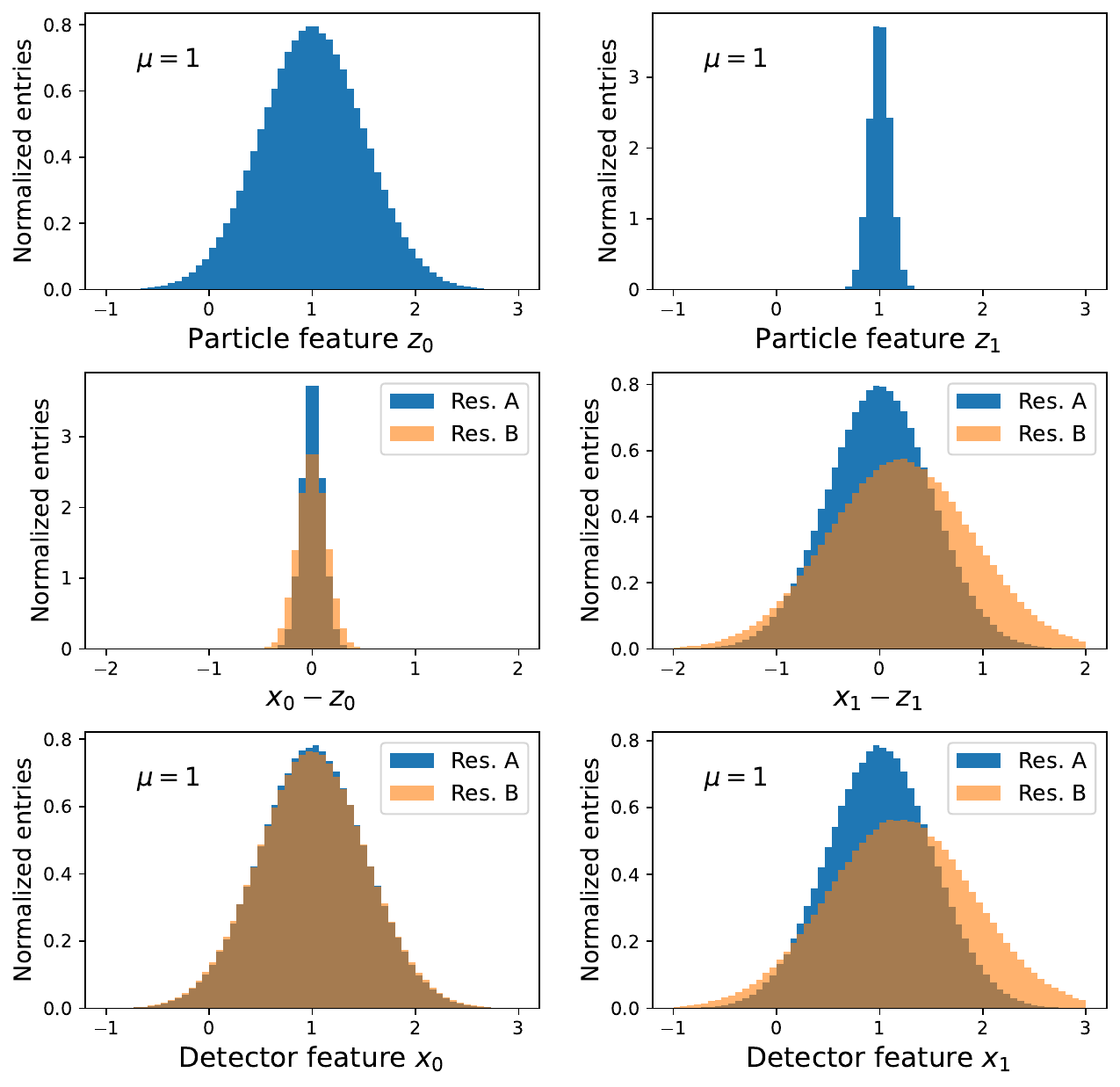}
   \caption{
      Input features and resolution model for toy regression example.
      Two different experimental resolution functions, A and B, are shown.
   }
   \label{fig:toy-regression-features}
   \end{center}
\end{figure}

Figure~\ref{fig:toy-regression-features} illustrates the input features and resolution model for the toy study.
Two particle-level features $z_0$ and $z_1$ are modeled as normal distributions: $Z_0\sim \mathcal{N}(\mu,0.5)$ and $Z_1\sim \mathcal{N}(\mu,0.1)$, where feature 1 is significantly more sensitive to the parameter of interest $\mu$.
The experimental resolution on the features is given by $X_0\sim\mathcal{N}(Z_0, 0.1)$ and $X_1\sim\mathcal{N}(Z_1, 0.5)$ so that feature 0 is well measured, while feature 1 has a relatively poor resolution.
For this model, the net experimental sensitivity to $\mu$ is the same for both features, but feature 0 is much less sensitive to detector effects.
Our proposed method will take this into account in the training of the neural network.

\begin{figure}[h]
   \begin{center}
\includegraphics[width=0.95\linewidth]{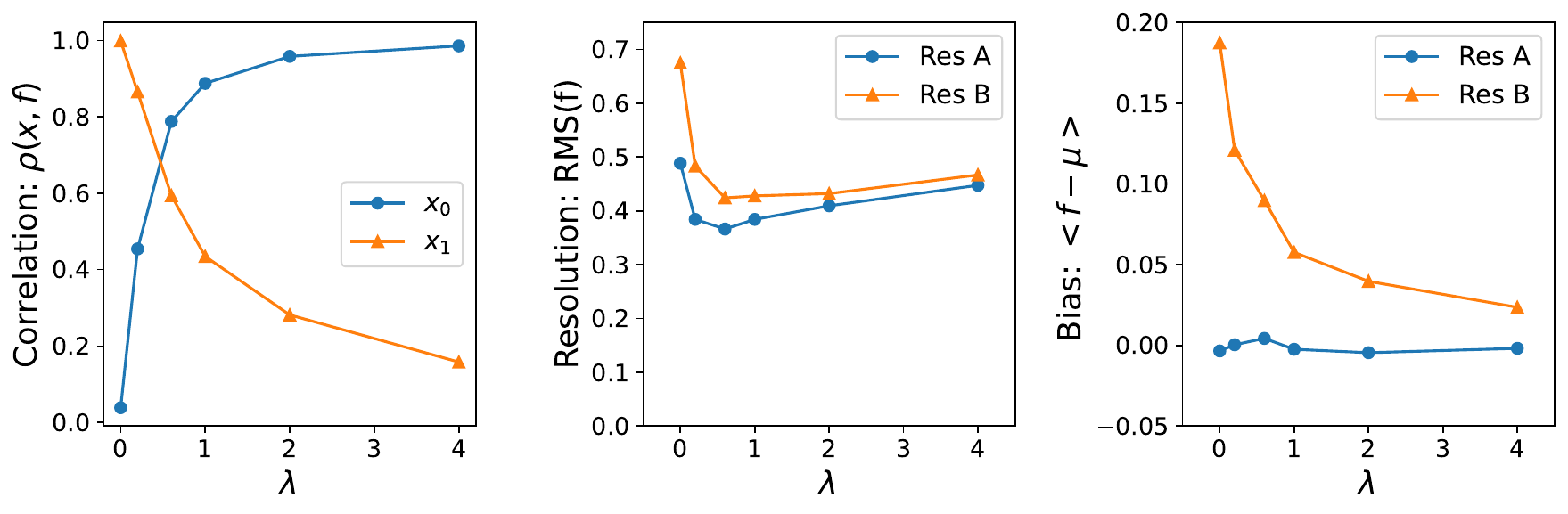}
   \caption{
       Results of toy regression example as a function of the $\lambda$ hyperparameter.
   }
   \label{fig:toy-regression-resolution}
   \end{center}
\end{figure}

To demonstrate the sensitivity to uncertainties associated with detector effects, we make predictions using $f$ trained with resolution model A on a sample generated with resolution model B, shown in Figure~\ref{fig:toy-regression-features}, where the width is increased by a factor of 1.4 for both features and a bias of 0.2 is introduced for the $x_1$ feature.
Figure~\ref{fig:toy-regression-resolution} shows the results as a function of the $\lambda$ hyperparameter.
With $\lambda=0$, the network relies almost entirely on feature 1, which has better particle-level resolution.
As $\lambda$ increases, more emphasis is placed on feature 0, which has much better detector-level resolution.
The resolution of the prediction starts at close to $\sqrt{0.5^2+0.1^2}$ for resolution model A, then reaches a minimum close to $\sqrt{0.5^2+0.1^2}/\sqrt{2}$ near $\lambda=0.5$ where both features have equal weight in the prediction.
The bias in the prediction for resolution model B is large for $\lambda=0$ but falls significantly with increasing $\lambda$.
This validates the key concept of the proposed method.

\subsection{ Full example for binary classification}
\label{sec:binarycase}

In the second example, we wish to design an observable that will discriminate between two different Monte Carlo parton shower models, while minimizing uncertainties from detector effects.  The objective will be to use the distribution of the observable to indicate which model best represents the data.  The discrimination test will be successful if the particle-level distribution of the observable for only one of the models is statistically consistent with the unfolded distribution of the observable from data.  Ideally, the difference in the shape of the particle-level distributions of the observable for the two models will be large compared to the size of the uncertainty from detector effects in the unfolding.

To design the observable for this task, we train a binary classification neural network to distinguish the two parton shower models, where we minimize detector effects with our additional term in the loss function.  The observable is trained to classify events, but this is not a case where each event in the data may be from one of two categories, such as in a neural network trained to discriminate signal from background.  All of the events in the data are from the same class.  The task is to use the shape of the observable to test which model is more consistent with the data.  The neural network is trained to find differences in the input features that allow it to distinguish the two models and this information is reflected in the shape of the neural network output distribution.  This makes the shape comparison the appropriate statistical test for the observable.

In the binary case, we have two datasets generated from simulation 1 (sim. 1) and simulation 2 (sim. 2).  
The loss function for classification is given by

\begin{align}
\label{eq:loss-classification}
    L[f] = -\sum_{i\in\text{sim. 1}} \log(f(z_i))-\sum_{i\in\text{sim. 2}} \log(1-f(z_i))+\lambda\sum_{i\in\text{sim. 1 \& 2}} (f(x_i)-f(z_i))^2\,,
\end{align}
where the first two terms represent the usual binary cross entropy loss function for classification and the third term represents the usual mean squared error loss term for regression tasks.  The notation $i\in\text{sim. j}$ means that $(z_i,x_i)$ are drawn from the $j^\text{th}$ simulation dataset.
As in the regression case, the hyperparameter $\lambda$ must be tuned and controls the trade off between sensitivity to the dataset and sensitivity to detector effects.  The network model is the same as in the previous example except that the final layer uses a sigmoid activation function.  The binary case is a special case of the previous section where there are only two values of the parameter of interest.  It may also be effective to train the binary case for a continuous parameter using two extreme values of the parameter.  In this paper, we use high-quality, well-curated datasets from the binary case because of their availability, but it would be interesting to explore the continuous case in the future.

To determine the efficacy of the new observable, we unfold the pseudodata.  Unfolding corrects for detector effects by performing regularized matrix inversion on the response matrix. We employ the widely-used \textsc{TUnfold} method~\cite{Schmitt:2012kp}, which is a least-squared-based fit with additional regularization.  We use \textsc{TUnfold} version 17.9~\cite{Schmitt:2012kp} through the interface included in the \textsc{Root} 6.24~\cite{Brun:1997pa} distribution.  
The response matrix is defined from a 2D binning the NN output, given detector level and particle level features.  The matrix uses 24 and 12 bins for detector and particle inputs, respectively, which gives reasonable stability and cross-bin correlations in the unfolding results.
The ultimate test is to show that the difference between sim.~1 at detector-level unfolded with sim.~2 for the response matrix and the particle level sim.~1 (or vice versa) is smaller than the difference between sim.~1 and sim.~2 at particle level.  In other words, this test shows that the ability to distinguish sim.~1 and sim.~2 significantly exceeds the modeling uncertainty from the unfolding.


\section{Datasets}
\label{sec:data}

We use deep inelastic scattering events from high-energy electron-proton collisions to demonstrate the performance of the new approach.  These simulated data are from the H1 experiment at HERA~\cite{H1:1996jzy,H1:1996prr} and are used in the same way as Ref.~\cite{Arratia:2021tsq}.  They are briefly described in the following.

Two parton shower Monte Carlo programs provide the particle-level simulation: \textsc{Rapgap}~3.1~\cite{Jung:1993gf}
or
\textsc{Djangoh}~1.4~\cite{Charchula:1994kf}.  The energies of the incoming beams are $E_e=27.6$ GeV and $E_p=920$ GeV, for the lepton and proton, respectively, matching the running conditions of HERA II.  Radiation from Quantum Electrodynamic processes is simulated by \textsc{Heracles} routines~\cite{Spiesberger:237380,Kwiatkowski:1990cx,Kwiatkowski:1990es}
in both cases.  The outgoing particles from these two datasets are then fed into a \textsc{Geant} 3~\cite{geant4}-based detector simulation.

Following the detector simulation, events are reconstructed with an energy-flow
algorithm~\cite{energyflowthesis,energyflowthesis2,energyflowthesis3} and the scattered electron is reconstructed using the default H1 approach~\cite{H1:2012qti,H1:2014cbm,H1:2021wkz}.  Mis-measured backgrounds are suppressed with standard selections~\cite{H1:2012qti,H1:2014cbm}.  This whole process makes use of the modernized H1 computing environment at DESY~\cite{Britzger:2021xcx}.  Each dataset is comprised of approximately 10 million events.

\begin{figure}[h]
   \begin{center}
\includegraphics[width=0.95\linewidth]{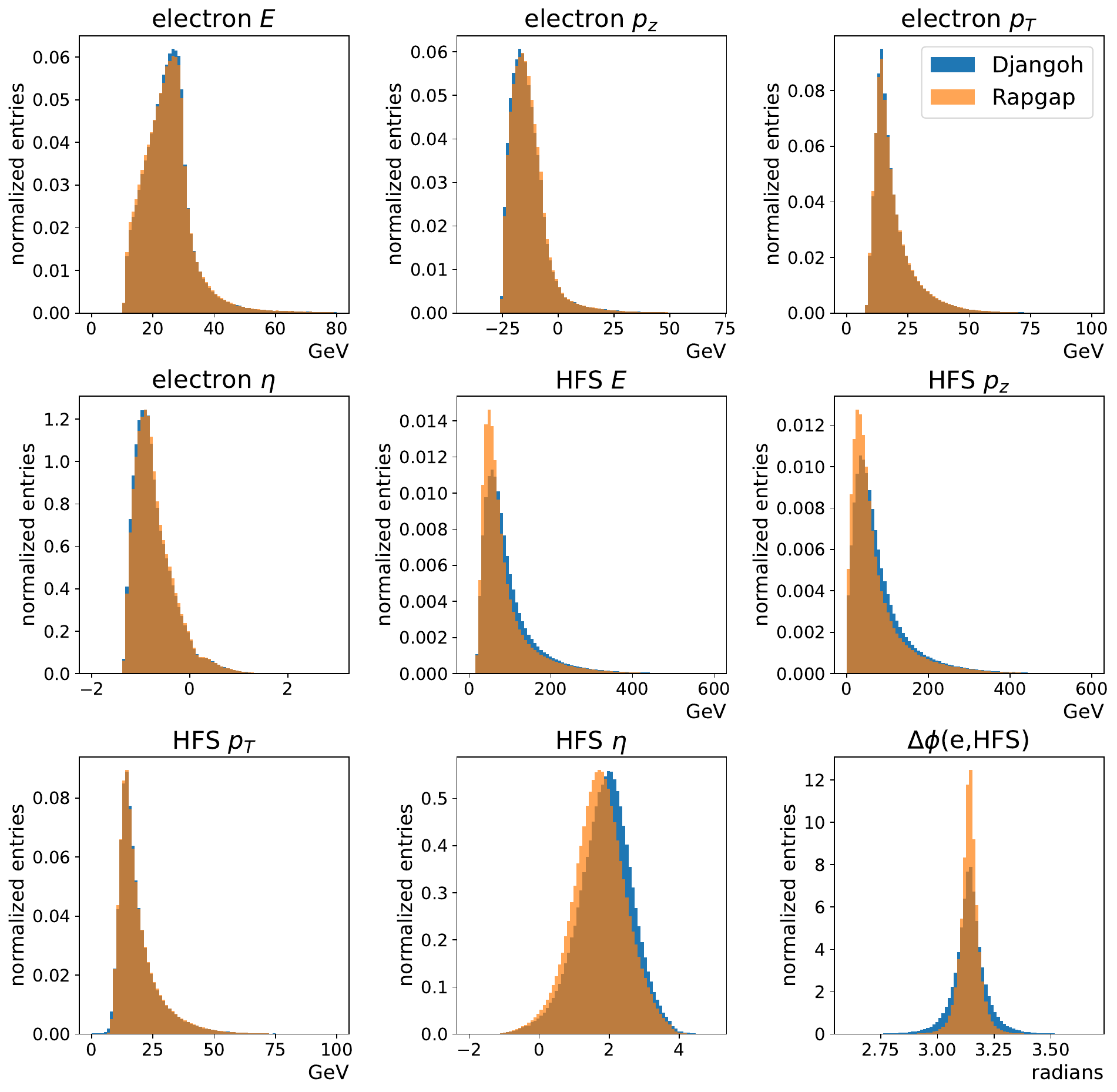}
   \caption{
      Particle level distributions of the nine NN input features for the \textsc{Djangoh} and \textsc{Rapgap} generators. 
   }
   \label{fig:feature-plots-particle-django-vs-rapgap}
   \end{center}
\end{figure}

Figure~\ref{fig:feature-plots-particle-django-vs-rapgap} shows histograms of the nine features used as input for the neural network training.  These features include the energy $E$, longitudinal momentum $p_z$, transverse momentum $p_T$, and pseudorapidity $\eta$ of the scattered electron and the total Hadronic Final State (HFS), as well as the difference in azimuthal angle between the two $\Delta \phi(e,{\rm HFS})$.  The HFS is quite sensitive to the $\eta$ acceptance of the detector.  In order to have HFS features that are comparable for the particle and detector definitions, we only use generated final state particles with $|\eta|<3.8$ in the definition of the particle-level HFS 4-vector.
Both simulations provide event weights that must be used for physics analysis.  In our study, we do not weight the simulated events in order to maximize the effective statistics of the samples in the neural network training.  This has a small effect on the spectra, but has a large impact on the number of effective events available for training.  The  electron feature distributions agree very well for the two simulations, while there are some visible differences in the HFS features.


\section{Results}
\label{sec:results}

\begin{figure}[ht!]
   \begin{center}
\includegraphics[width=0.95\linewidth]{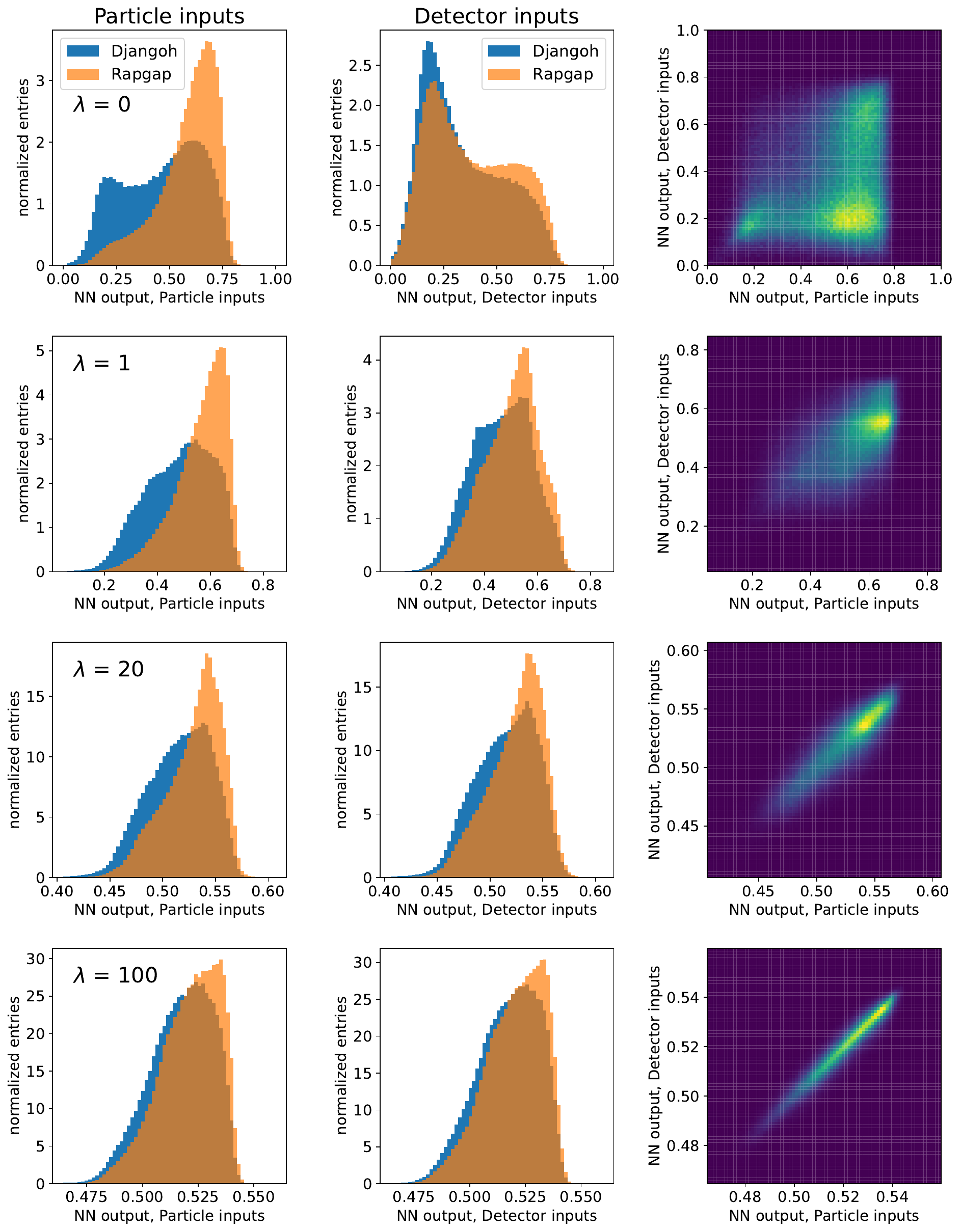}
   \caption{
      Neural Network output distributions for four values of the $\lambda$ hyperparameter, which sets the scale for the Detector - Particle disagreement penalty in the loss function.
      The top row shows the results for $\lambda = 0$, where there is no penalty if the NN predictions for Detector-level input features and Particle-level input features disagree.
      The bottom three rows show increasing values of $\lambda$ : 1, 20, and 100.
   }
   \label{fig:training-results-varying-lambda}
   \end{center}
\end{figure}
%
We now apply the method introduced in Sec.~\ref{sec:binarycase} to the DIS dataset described in Sec.~\ref{sec:data}. Figure~\ref{fig:training-results-varying-lambda} shows the results of four neural network trainings with different values of $\lambda$, which sets the relative weight of the MSE term in the loss function (Eq.~\ref{eq:loss-classification}) that controls the sensitivity to detector effects.  With $\lambda=0$,
the classification performance for particle-level inputs is strong, while there are significant disagreements between the particle and detector level neural network outputs. 
As $\lambda$ increases, the particle and detector level agreement improves at the cost of weaker classification performance.  In what follows, we will use the network trained with $\lambda=100$.  For a parameter estimation task, the entire distribution will be used for inference and therefore excellent event-by-event classification is not required.

\begin{figure}
   \begin{center}
\includegraphics[width=0.31\linewidth]{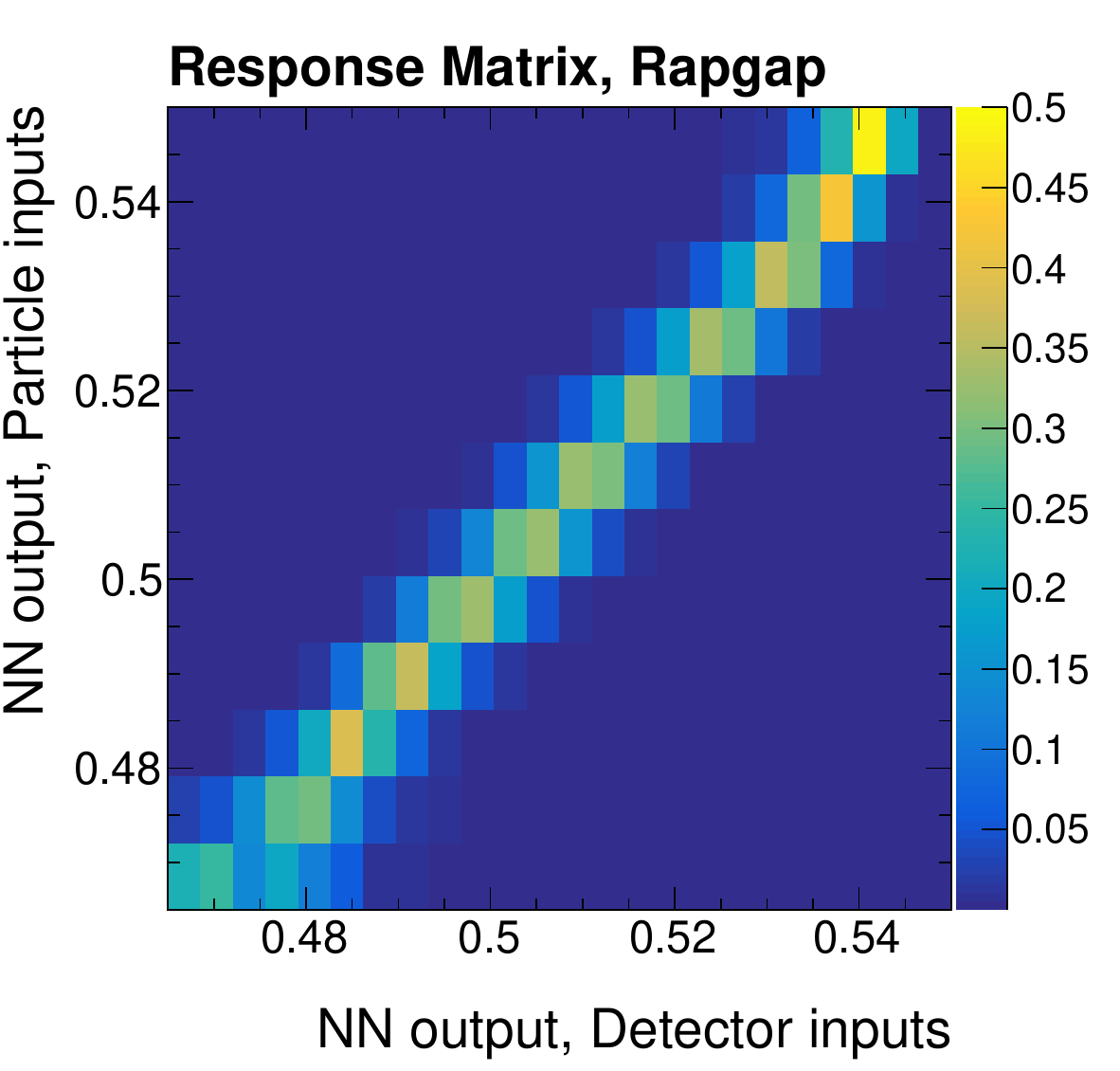}
\includegraphics[width=0.31\linewidth]{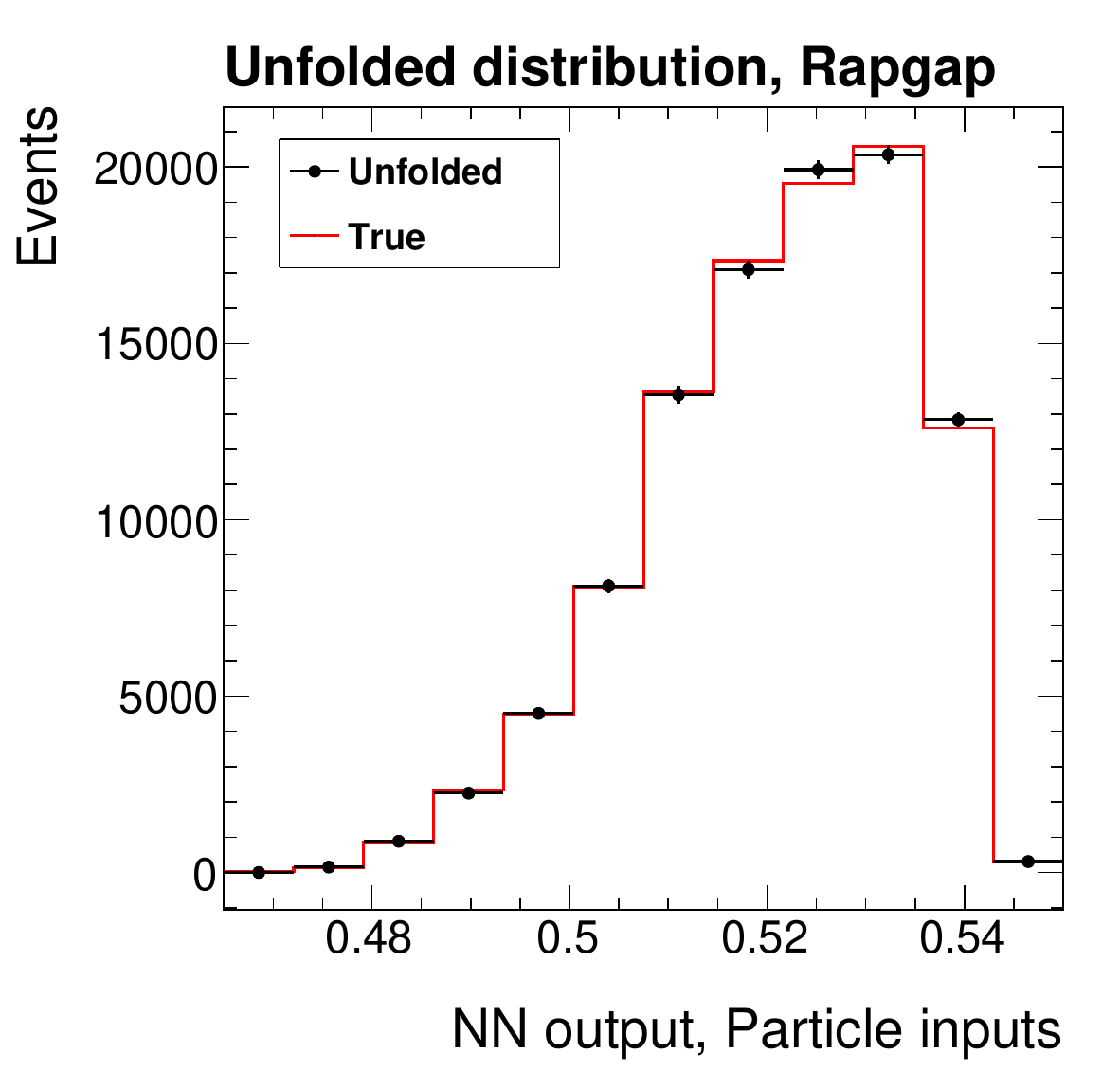}
\includegraphics[width=0.31\linewidth]{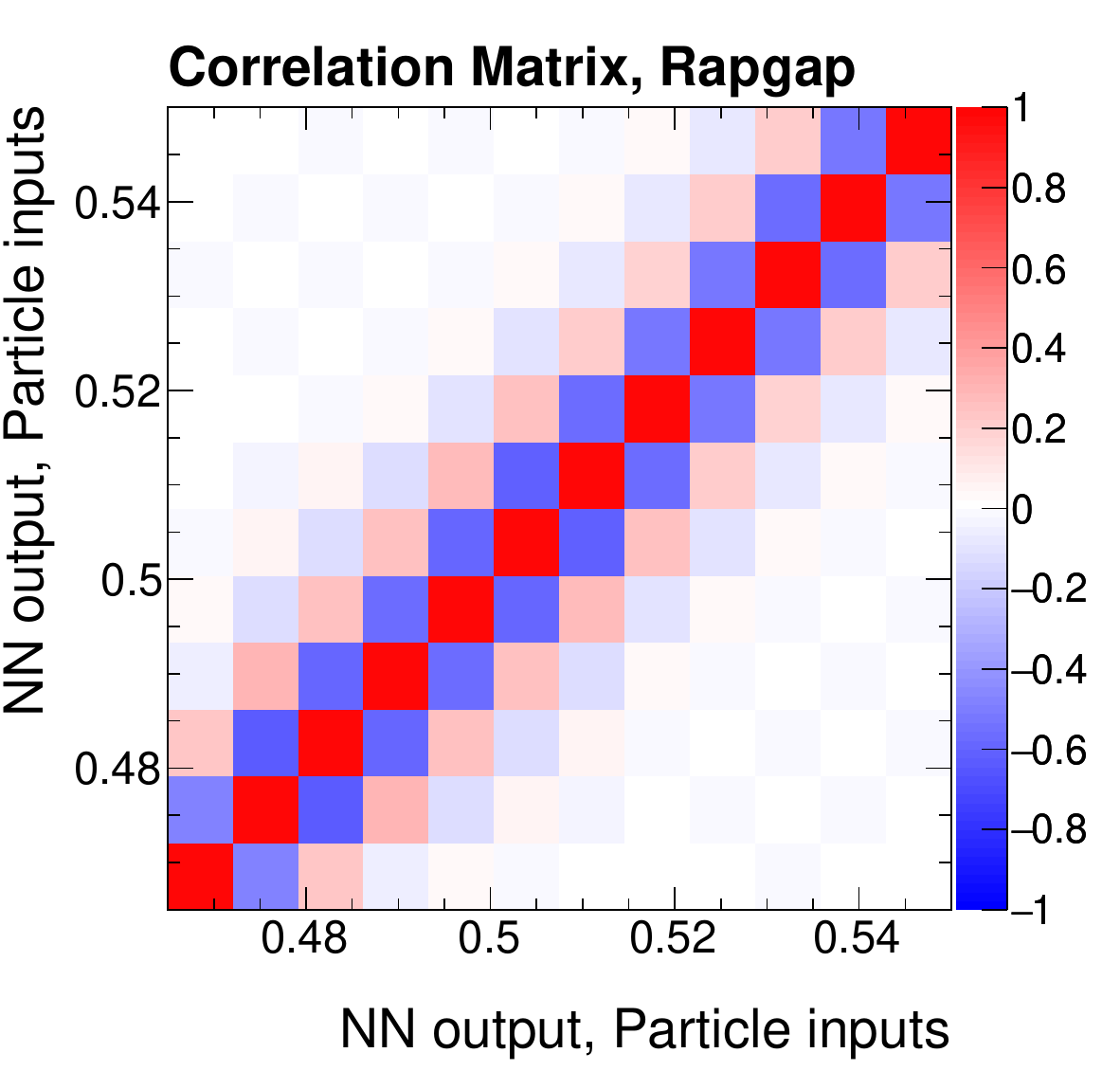} 
\includegraphics[width=0.31\linewidth]{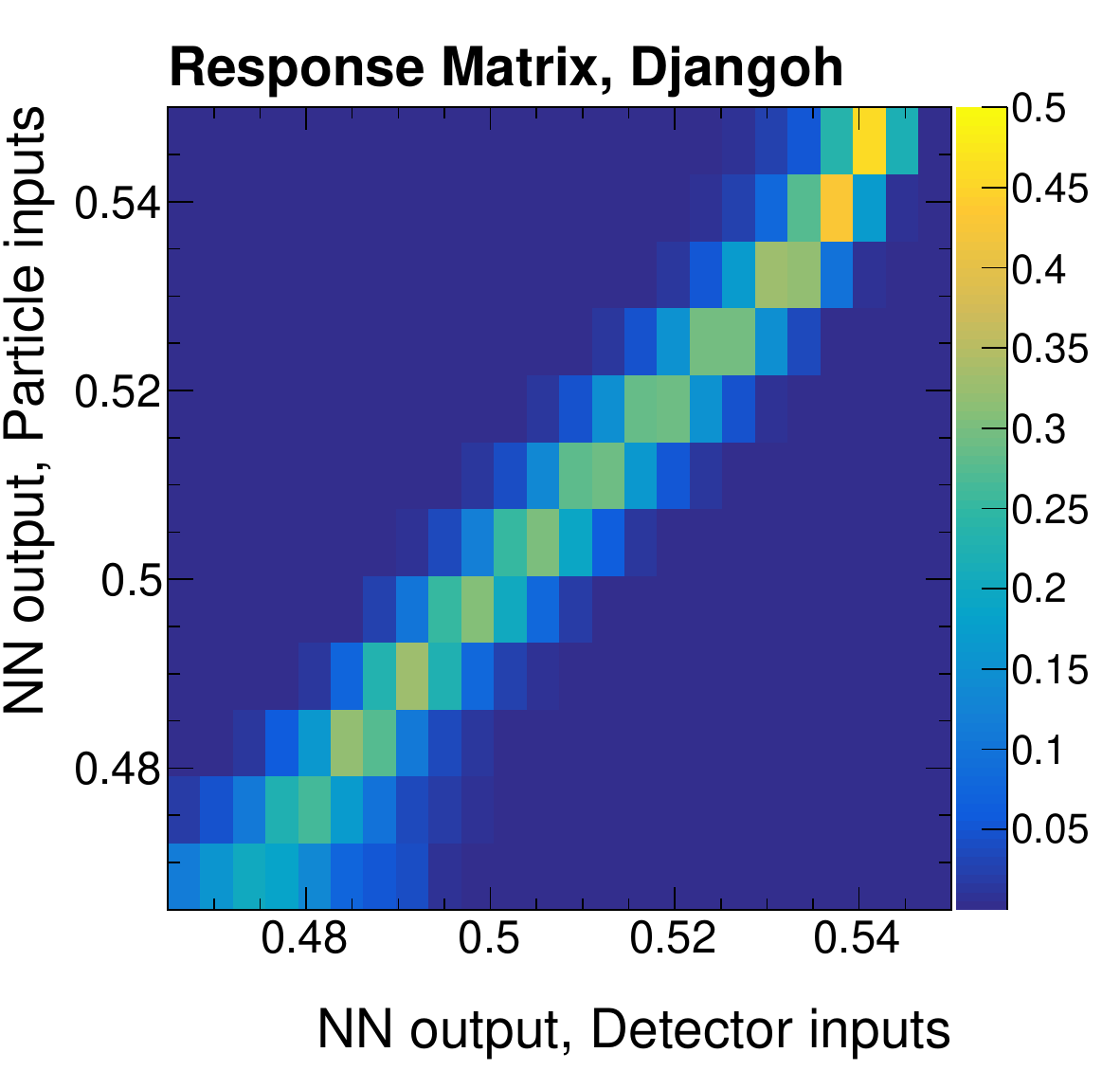}
\includegraphics[width=0.31\linewidth]{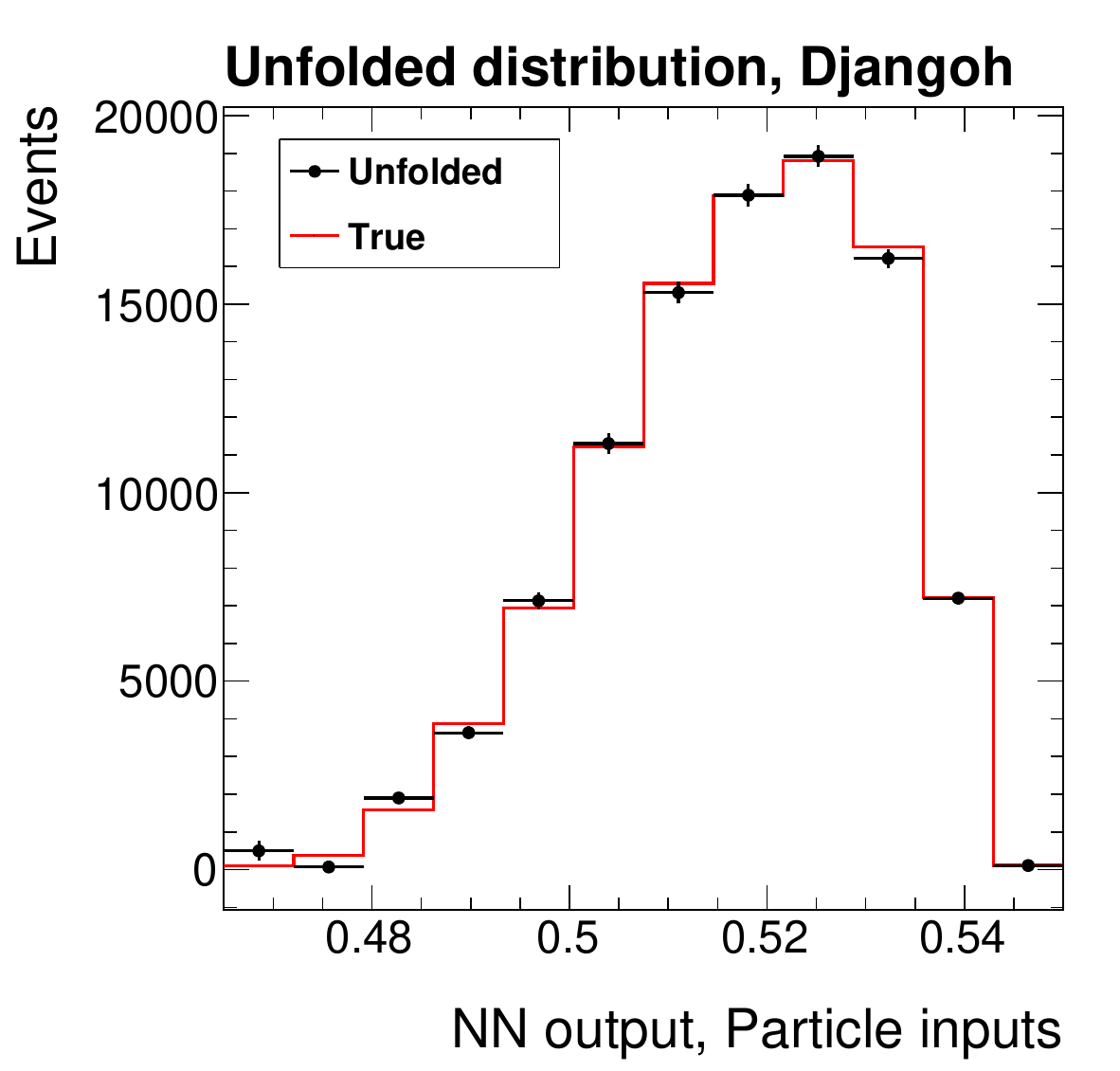}
\includegraphics[width=0.31\linewidth]{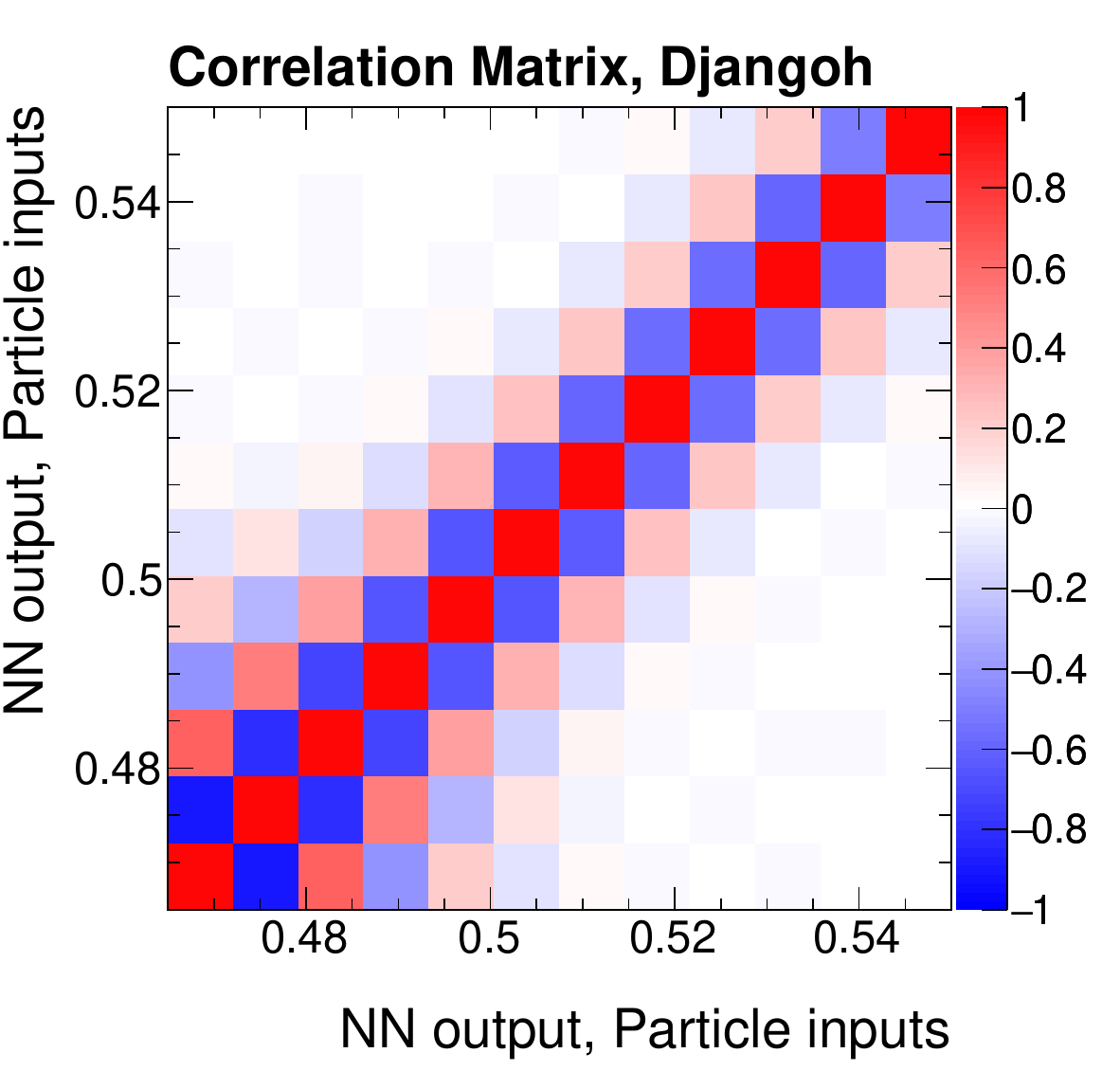} 
   \caption{
      Results of unfolding the NN output.  The top (bottom) row shows the unfolding for the \textsc{Rapgap} (\textsc{Djangoh}) generator.  
      The left column shows the response matrix for the unfolding, where the distribution of the NN output given the Detector-based input features (horizontal axis) is normalized to unit area for each bin of the NN output given the Particle-based input features (vertical axis).
      The center column shows the unfolded distribution compared to the true distribution.
      The right column shows the matrix of correlation coefficients from the unfolding.
   }
   \label{fig:unfolding-results}
   \end{center}
\end{figure}
%

Next, we investigate how these spectra are preserved after unfolding.  Figure~\ref{fig:unfolding-results} shows the results of unfolding the neural network output, given detector-level features, to give the neural network output distribution for particle-level features.  The input distribution for the unfolding is $10^5$ events randomly chosen from a histogram of the neural network output for detector-level inputs from the simulation.  The unfolding response matrices for the two simulations agree fairly well and are concentrated along the diagonal.  The output of the unfolding shows very good agreement with the true distribution of the neural network output given particle-level inputs, demonstrating acceptable closure for the unfolding.  The correlations in the unfolding result are mostly between neighboring bins of the distribution.

\begin{figure}[h]
   \begin{center}
\includegraphics[width=0.45\linewidth]{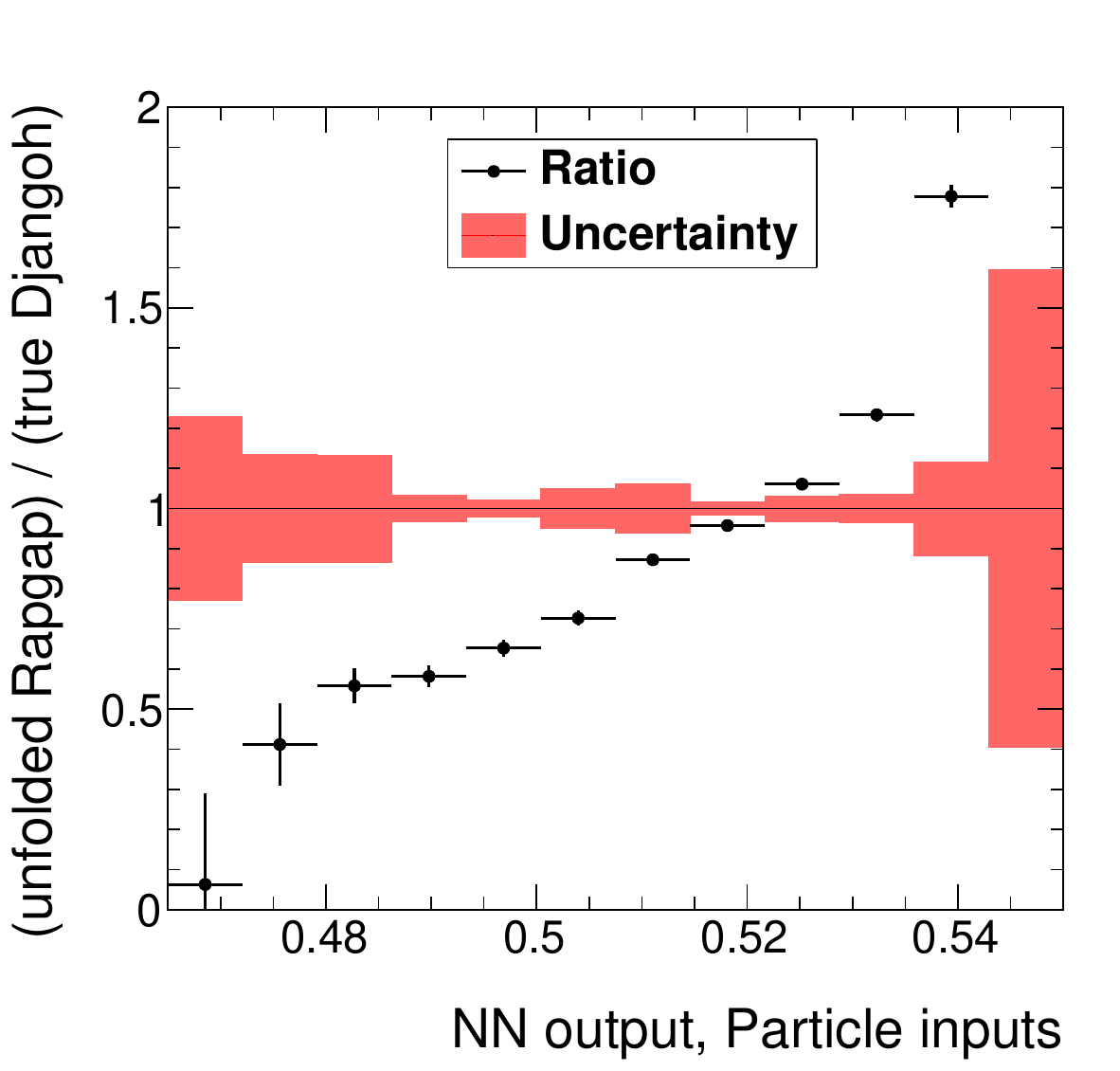}
\includegraphics[width=0.45\linewidth]{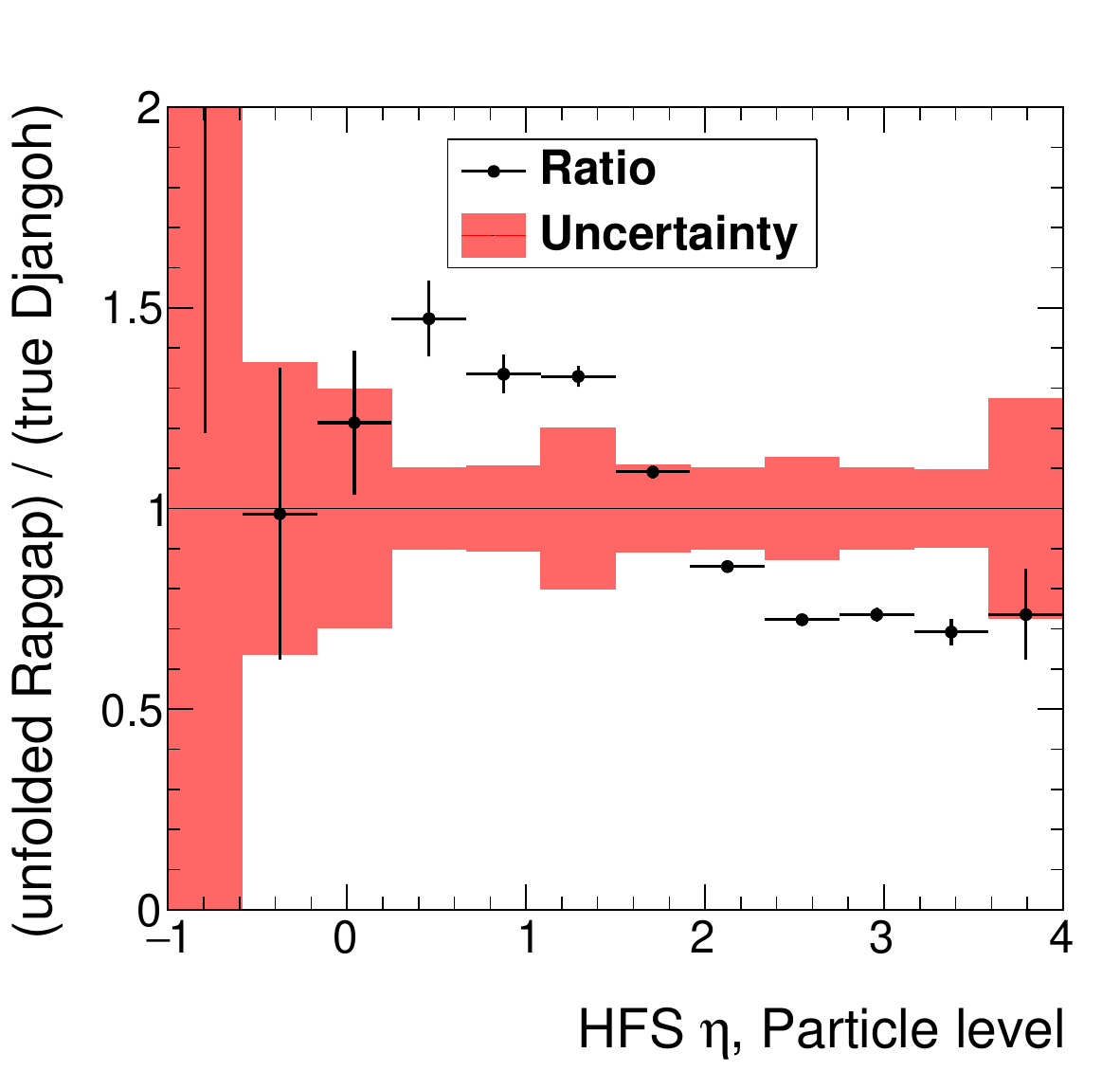}

   \caption{
      Model discrimination sensitivity and uncertainty for the neural network output distribution (left) and the hadronic final state $\eta$ distribution (right).
      The uncertainty, shown by the shaded red distribution, is the model dependence of the unfolding added in quadrature with the statistical error from the unfolding (error bars on the black points).
      The black points show the ratio of the unfolded distribution from the \textsc{Rapgap} simulation divided by the particle-level distribution from the \textsc{Djangoh} simulation.  
      The significance of the deviation from unity is a measure of the model discrimination sensitivity.
      Additional distributions and comparisons are given in Appendix~\ref{app:model-dependence-details}.
   }
   \label{fig:gen-syst-model-disc-both}
   \end{center}
\end{figure}
%

One of the biggest challenges for the $\lambda=0$ case is that it is highly sensitive to regions of phase space that are not well-constrained by the detector.  As a result, the output of the unfolding is highly dependent on the simulation used in the unfolding (prior). 
The left side of Figure~\ref{fig:gen-syst-model-disc-both} shows the model dependence of the unfolding and the ability of the neural network to perform the model classification task.
The unfolding model dependence is estimated from a comparison with using the response matrix from the other simulation.
We test the model discrimination sensitivity by dividing the unfolded distribution by the particle-level distribution from the other simulation.
The degree to which this ratio deviates from unity is a measure of the model discrimination power of the method.
For the neural network, the model dependence of the unfolding is small and generally less than 10\%.
The size of the deviation from unity for the neural network is large compared to the size of the uncertainty, which is dominated by the model dependence, indicating that the neural network can distinguish the two simulations.

Figure~\ref{fig:feature-plots-particle-django-vs-rapgap} shows that some of the HFS variables in the input features may be able to distinguish the two models directly.
The right side of Figure~\ref{fig:gen-syst-model-disc-both} shows the results of running the unfolding procedure using the HFS $\eta$ distribution for model discrimination, where the model dependence is significantly larger,  compared to the neural network output.
The shape is distorted, including deviations up to 20\%, when the response matrix from the other simulation was used in the unfolding.  Since the modeling uncertainty is comparable to or larger than the size of the effect we are trying to probe, such observables are much less useful than the neural network output for the inference task.

We perform a quantitative evaluation of the model discrimination power for this example using a $\chi^2$ computed from the difference between the unfolded distribution and the particle-level distribution for a given model.  The uncertainties in the $\chi^2$ are from a combination of the unfolding covariance matrix and a covariance matrix for model dependence uncertainty from the unfolding response matrix.  The distribution of the $\chi^2$ from a set of toy Monte Carlo experiments with 2000 events per experiment is close to that from a $\chi^2$ PDF with 12 degrees of freedom when the model for the comparison in the $\chi^2$ is the same as the model used to generate the toy samples (\textsc{Djangoh}), which validates the unfolding statistical uncertainty in the covariance matrix and the $\chi^2$ calculation.  We use this distribution to define a $\chi^2$ threshold for the critical region corresponding to a frequency of 1\% for the correct model hypothesis to have a $\chi^2$ greater than the threshold.  When the toy samples are drawn from the alternative model (\textsc{Rapgap}), we find that the frequency for the $\chi^2$ to be above the threshold is 98.6\% for the designer neural network observable, but only 63.0\% for the HFS $\eta$ observable.  This shows that the designer neural network observable has superior discrimination power.


\section{Conclusions and Outlook}
\label{sec:conclusion}

Unfolded differential cross section measurements are a standard approach to making data available for downstream inference tasks. While some measurements can be used for a variety of tasks, often, there is a single goal that motivates the result.  In these cases, we advocate to design observables that are tailored to the physics goal using machine learning.  The output of a neural network trained specifically for the downstream task is an observable and its differential cross section likely contains more information than classical observables.  We have proposed a new loss function for training the network so that the resulting observable can be well measured.  The neural network observable is thus trained using a loss function composed of two parts: one part that regresses the inputs onto a parameter of interest and a second part that penalizes the network for producing different answers at particle level and detector level.  A tunable, and problem-specific hyperparameter determines the tradeoff between these two goals.  We have demonstrated this approach with both a toy and physics example.  For the deep inelastic scattering example, the new approach is shown to be much more sensitive than classical observables while also having a reduced dependence on the starting simulation used in the unfolding.  We anticipate that our new approach could be useful for a variety of scientific goals, including measurements of fundamental parameters like the top quark mass and tuning Monte Carlo event generators.

There are a number of ways this approach could be extended in the future.  We require that the observable have the same definition at particle level and detector level, while additional information at detector-level like resolutions may be useful to improve precision.  A complementary strategy would be to use all the available information to unfold the full phase space~\cite{Arratia:2021otl}.  Such techniques may improve the precision by integrating all of the relevant information at detector level, but they may compromise specific sensitivity by being broad and have no direct constraints on measurability.  It would be interesting to compare our tailored approach to full phase space methods in the future.


\section*{Code availability}
The code in this work can be found in: \url{https://github.com/owen234/designer-obs-paper}.

\section*{Acknowledgments}

We thank Miguel Arratia and Daniel Britzger for useful discussions and feedback on the manuscript.  Additionally, we thank our colleagues from the H1 Collaboration for allowing us to use the
simulated MC event samples. Thanks to DESY-IT and the MPI f{\"u}r Physik for providing some computing infrastructure and supporting the data preservation project of the HERA experiments.
B.N. was supported by the Department of Energy, Office of Science under contract number DE-AC02-05CH11231.

\newpage
\appendix
\section{ Alternative loss function for the regression example}
\label{alt-loss}

\begin{figure}[h]
   \begin{center}
\includegraphics[width=0.95\linewidth]{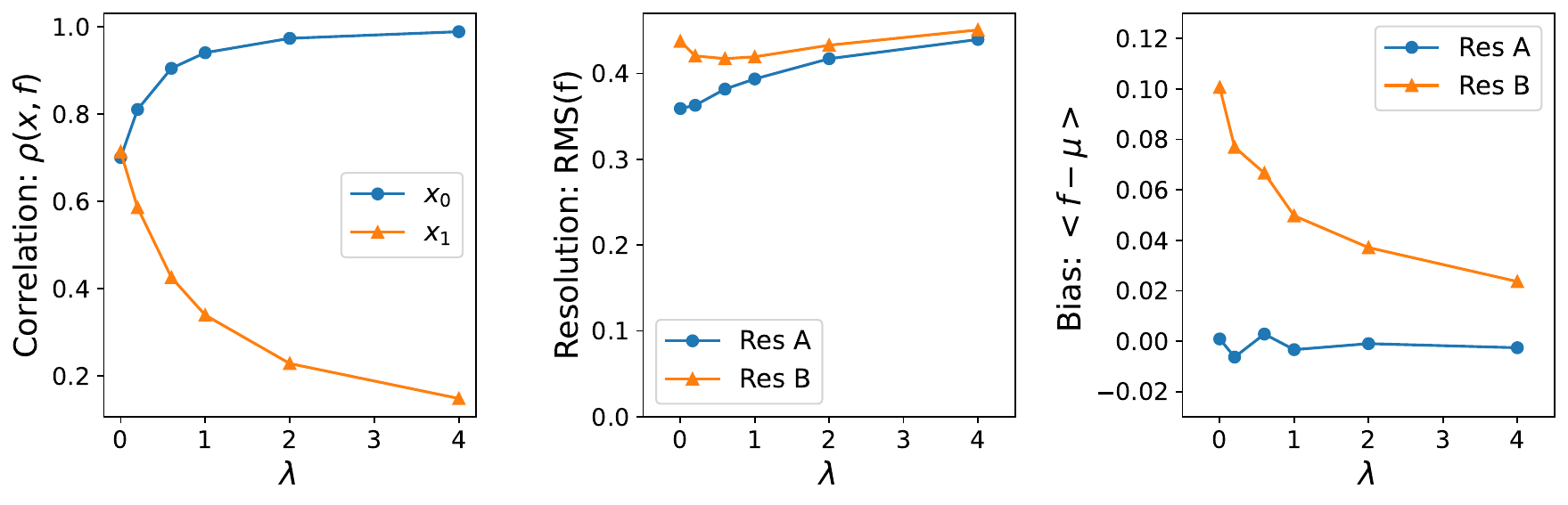}
   \caption{
       Results of toy regression example as a function of the $\lambda$ hyperparameter, using Equation~\ref{eqn:alt-loss} for the loss function in the  training.
   }
   \label{fig:toy-regression-resolution-alternative-loss}
   \end{center}
\end{figure}

An alternative approach in the regression example is to use the detector-level features instead of the particle-level features in the first term of the loss function.
Figure~\ref{fig:toy-regression-resolution-alternative-loss} shows the results of using Equation~\ref{eqn:alt-loss} instead of Equation~\ref{eq:loss-regression} in the training.
\begin{align}
\label{eqn:alt-loss}
    L[f] = \sum_{i} (f(x_i) - \mu_i)^2
    + \lambda\sum_{i} (f(x_i)-f(z_i))^2\,,
\end{align}
%
With $\lambda=0$, which corresponds to the usual approach for this type of regression task, the correlation between each of the detector level features $x_0$ and $x_1$ and the network prediction $f$ is the same, reflecting the fact that $x_0$ and $x_1$ have the same sensitivity to $\mu$.
As $\lambda$ increases, more emphasis is placed on feature $x_0$, which is well measured.
The resolution of the regression, given by the RMS of $f$, starts at the expected value of about $\sqrt{0.5^2+0.1^2}/\sqrt{2}$ for resolution model A and $\lambda=0$ and increases with $\lambda$ as the network relies more on $x_0$ for the prediction.

\section{Additional distributions for the model dependence and discrimination tests}
\label{app:model-dependence-details}

\begin{figure}[h]
   \begin{center}
\includegraphics[width=0.40\linewidth]{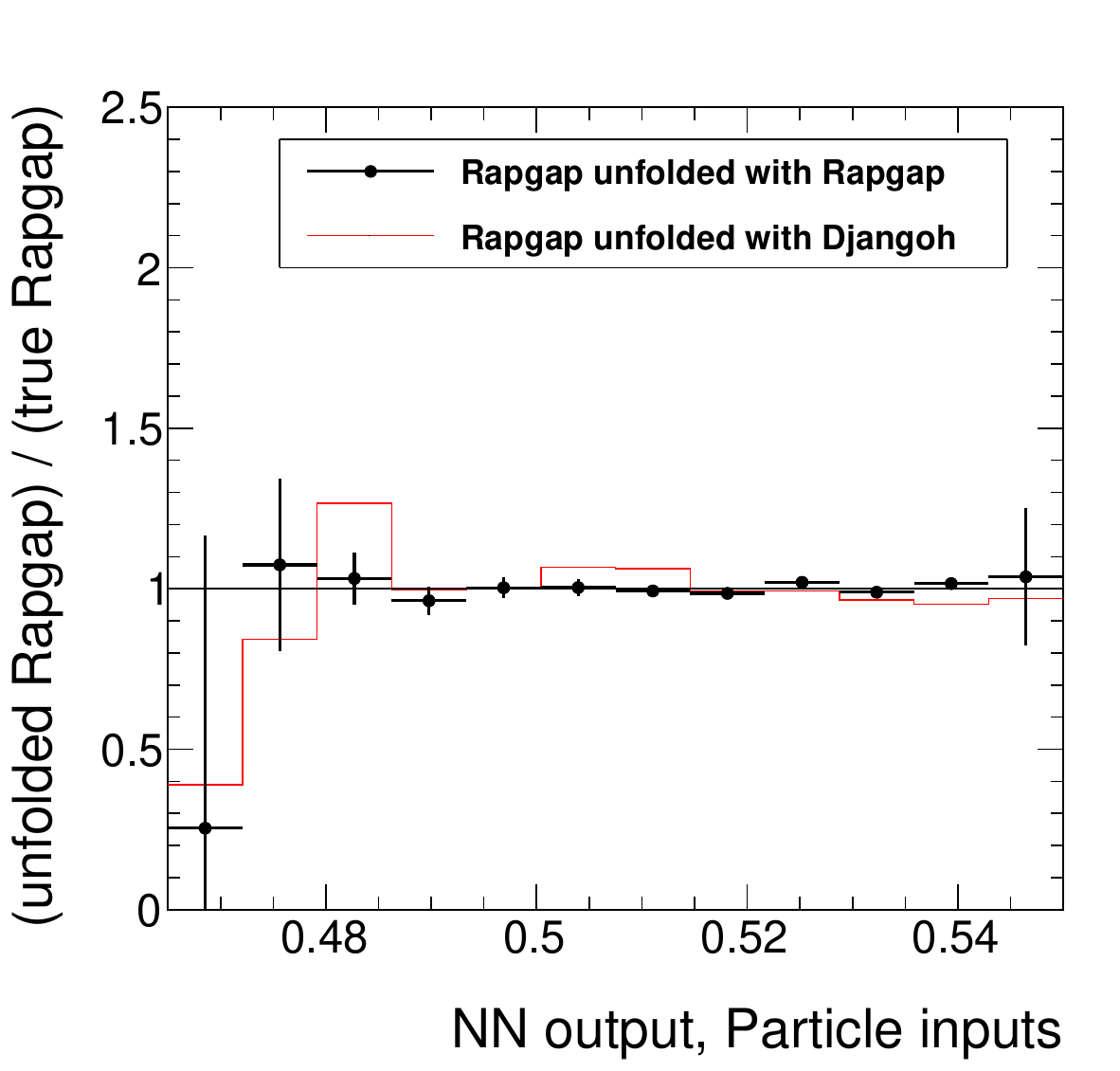}
\includegraphics[width=0.40\linewidth]{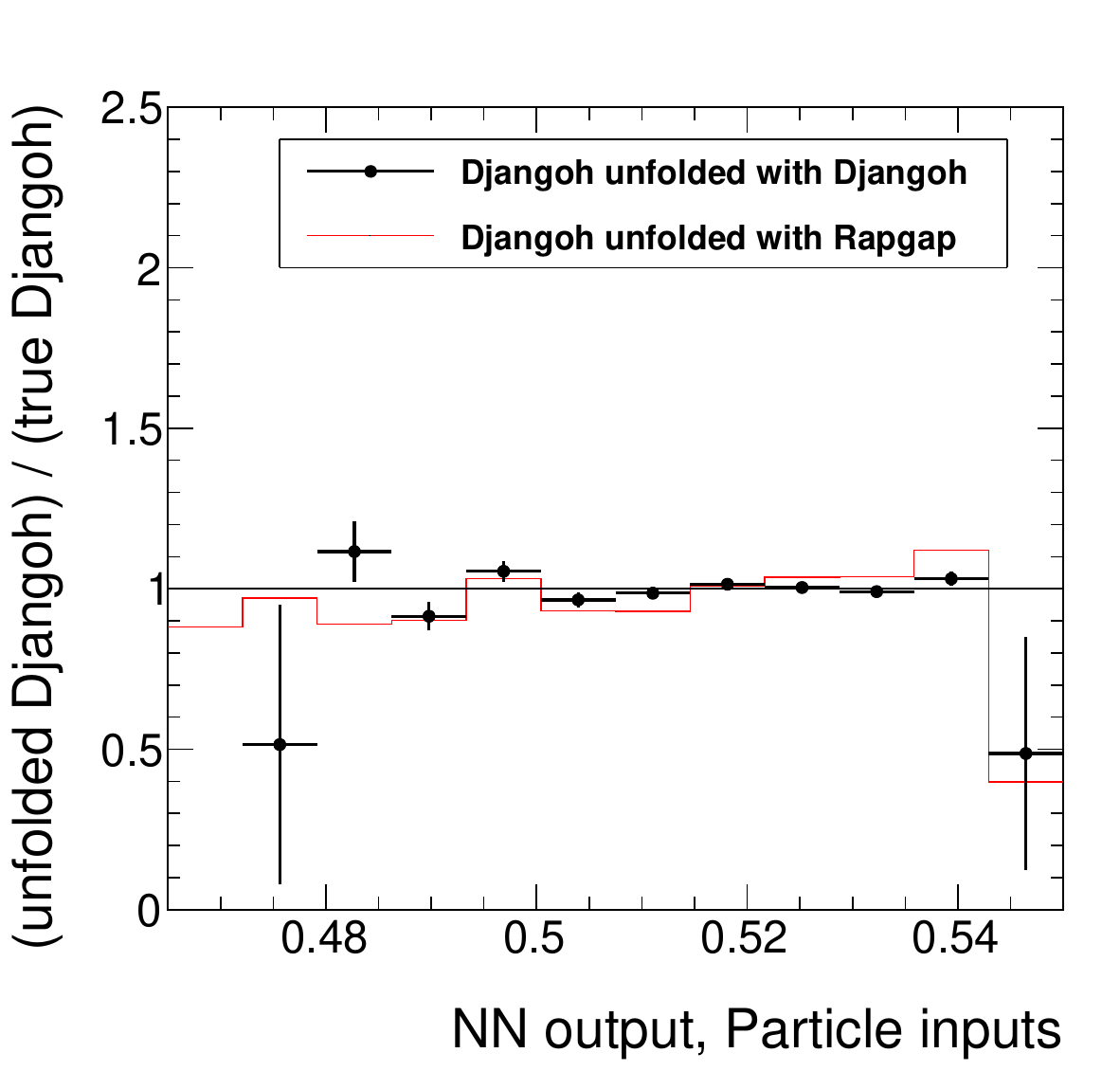}
\includegraphics[width=0.40\linewidth]{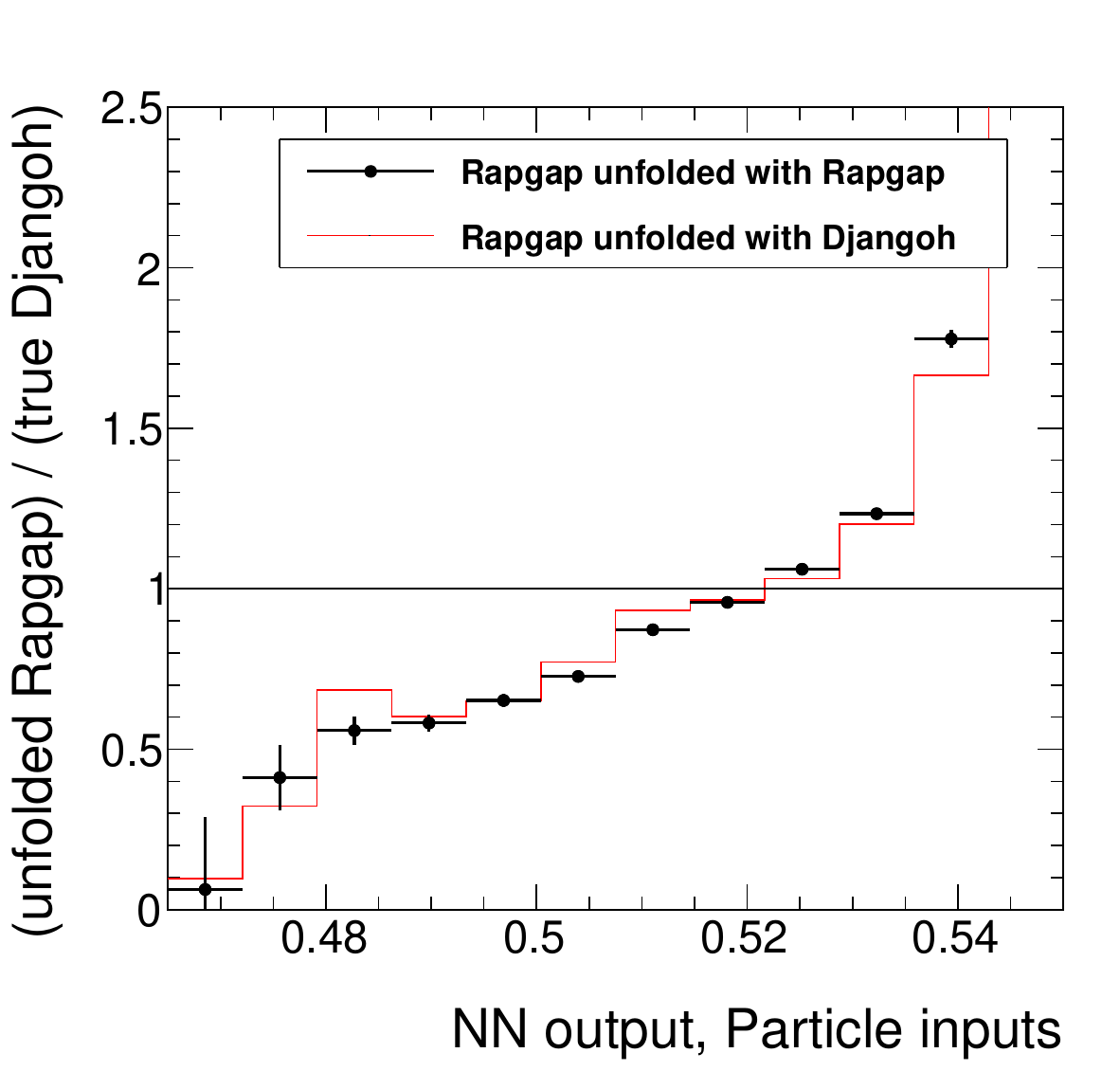}
\includegraphics[width=0.40\linewidth]{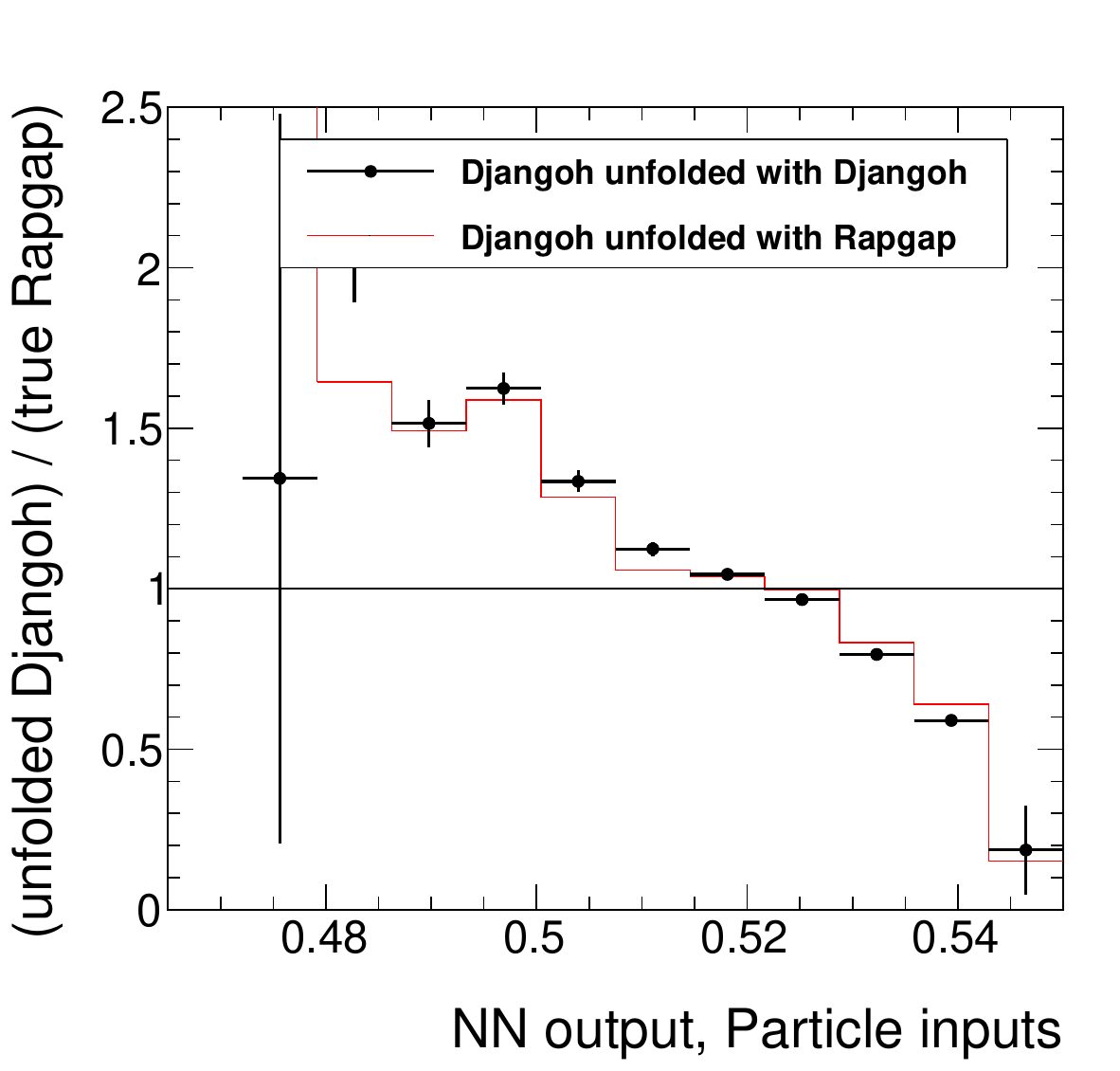}

   \caption{
      Results of testing the model dependence of the unfolding the {\bf neural network output} distribution by varying the response matrix used in the unfolding.
      The top row normalizes the unfolded distribution with the true distribution from the same simulation, testing the unfolding closure (points) and unfolding model dependence (red histogram).
      The bottom row normalizes the unfolded distribution with the true distribution from the other simulation, showing the ability to distinguish the models.
   }
   \label{fig:gen-syst-model-disc-nnsf3}
   \end{center}
\end{figure}

\begin{figure}[h]
   \begin{center}
\includegraphics[width=0.40\linewidth]{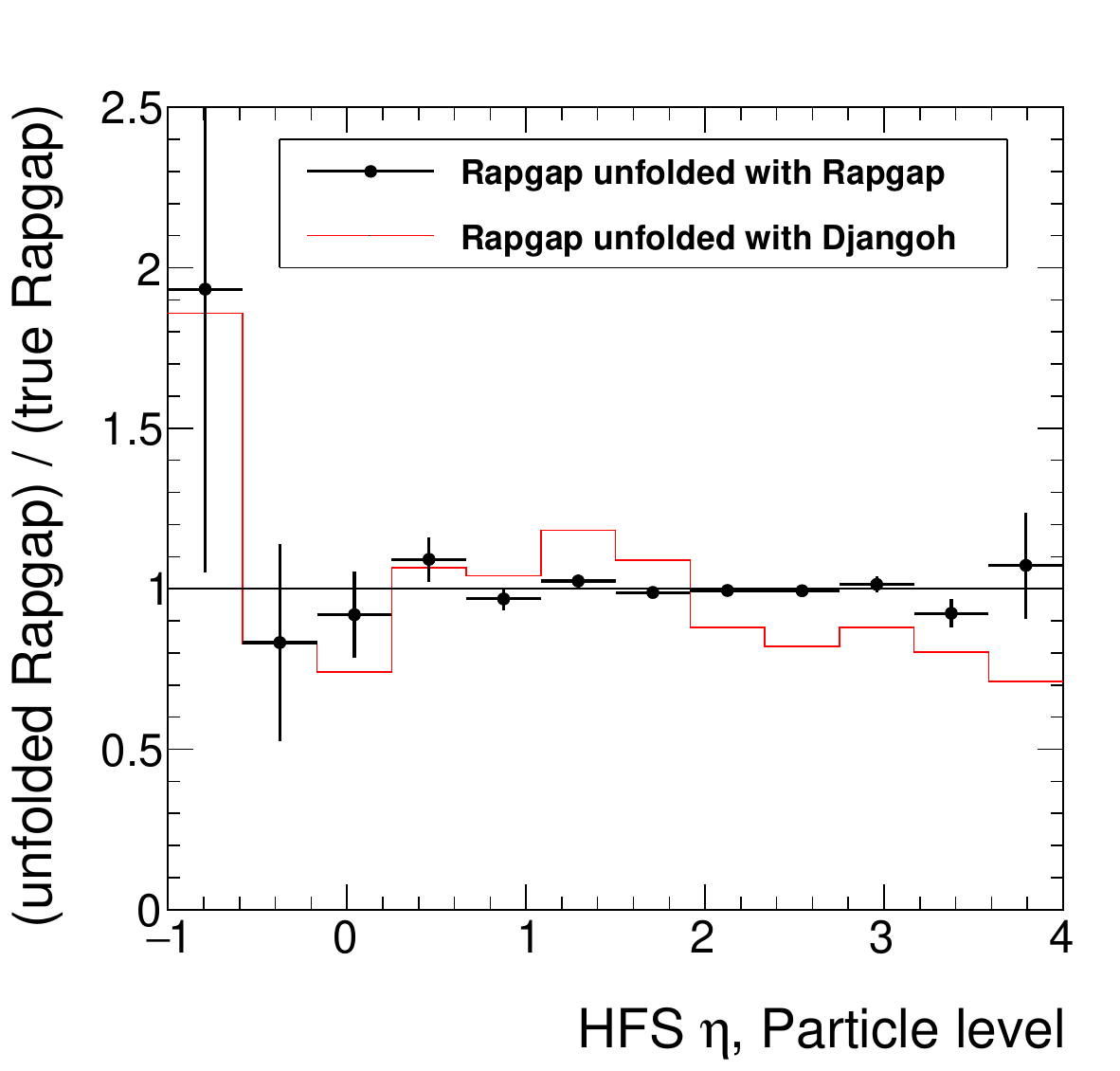}
\includegraphics[width=0.40\linewidth]{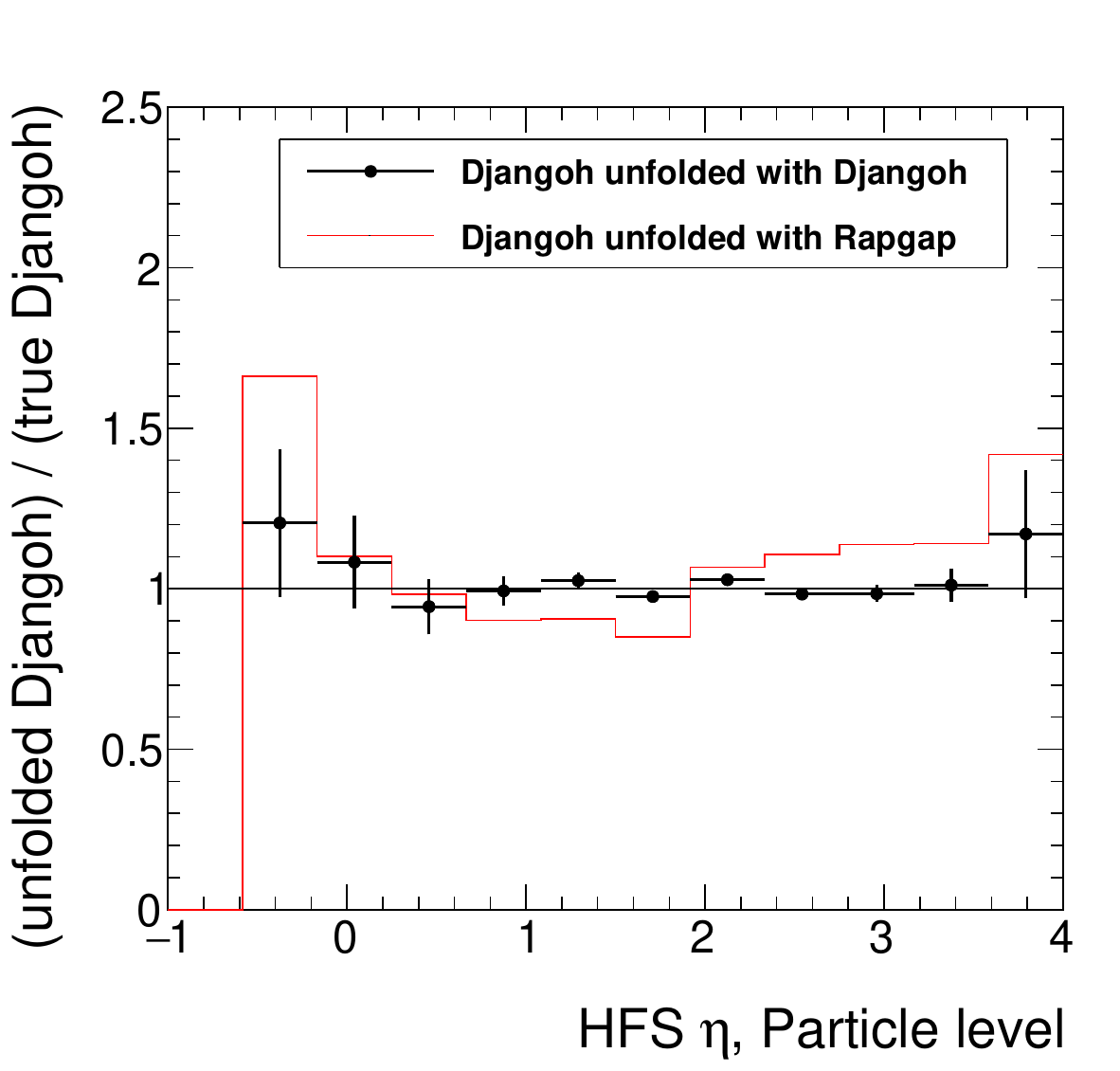}
\includegraphics[width=0.40\linewidth]{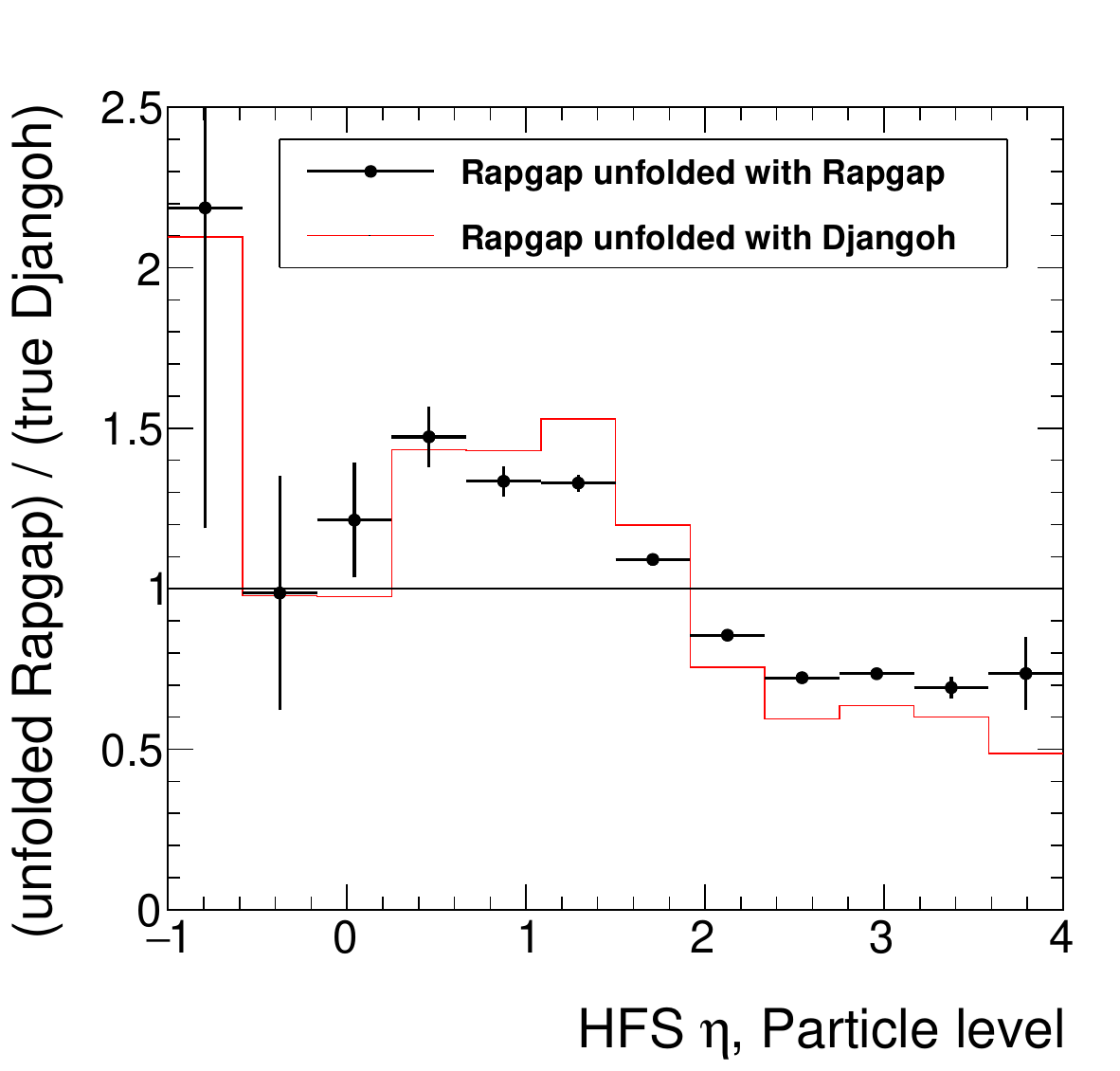}
\includegraphics[width=0.40\linewidth]{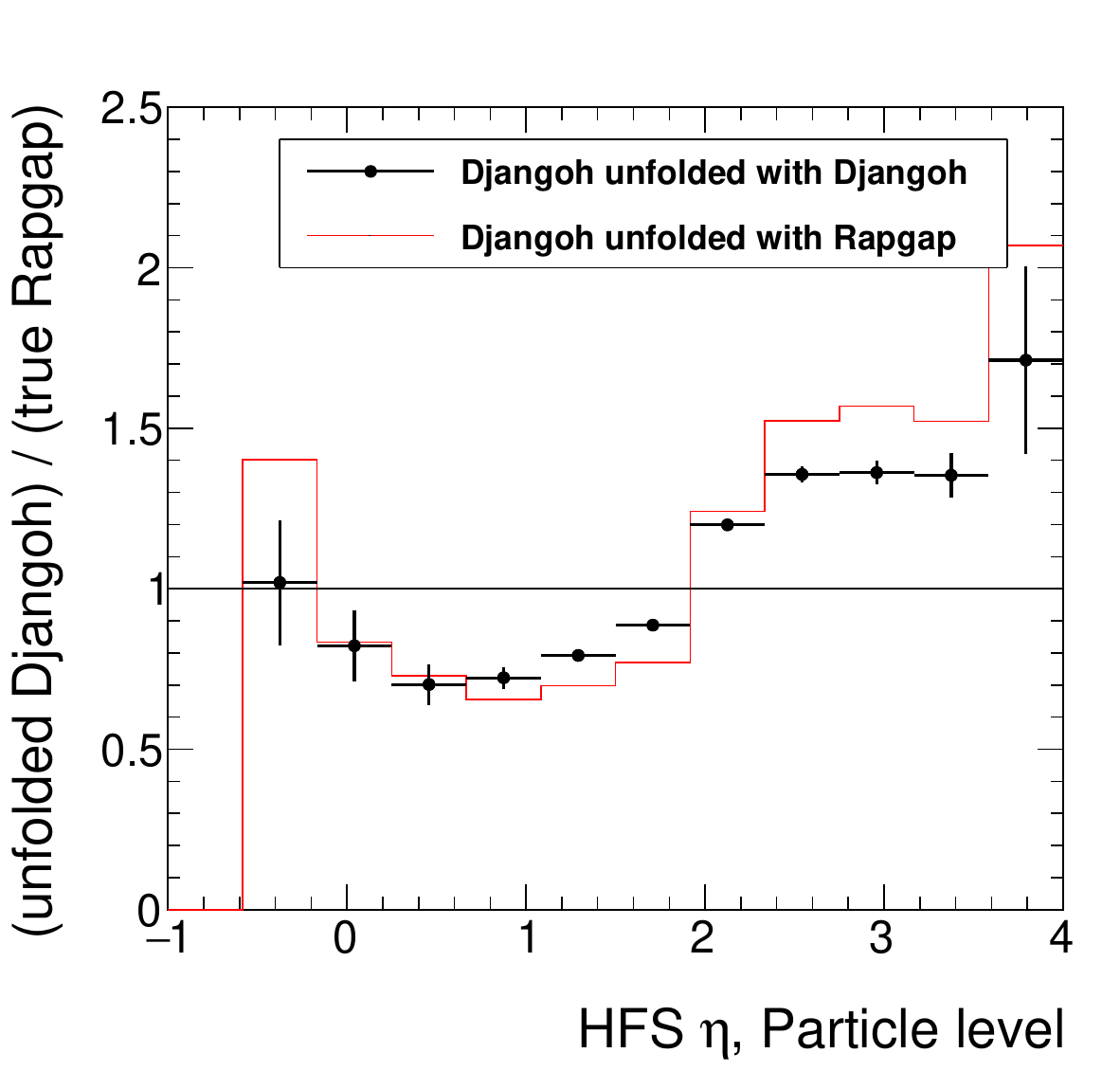}

   \caption{
      Results of testing the model dependence of the unfolding the {\bf HFS {\boldmath $\eta$}} distribution by varying the response matrix used in the unfolding.
      The top row normalizes the unfolded distribution with the true distribution from the same simulation, testing the unfolding closure (points) and unfolding model dependence (red histogram).
      The bottom row normalizes the unfolded distribution with the true distribution from the other simulation, showing the ability to distinguish the models.
   }
   \label{fig:gen-syst-model-disc-hfseta}
   \end{center}
\end{figure}

Figures~\ref{fig:gen-syst-model-disc-nnsf3} and~\ref{fig:gen-syst-model-disc-hfseta} show the model dependence of the unfolding and model discrimination sensitivity for the neural network and the HFS $\eta$ distributions, respectively.
The top row of each figure shows the results of the closure tests compared with a measure of the unfolding model dependence, where we perform the unfolding with the response matrix from the other simulation.
The unfolded distributions have been normalized using the true distribution from the same simulation, giving an expected flat distribution consistent with 1.
The bottom row shows the unfolded distributions instead normalized by the true distribution from the other simulation.
The degree to which the ratio deviates from unity is a measure of the  model discrimination power of the network.
Figure~\ref{fig:gen-syst-model-disc-both} in the main text uses the lower-left distribution from Figures~\ref{fig:gen-syst-model-disc-nnsf3} and~\ref{fig:gen-syst-model-disc-hfseta} to display the results.

The unfolding shows good closure for both the neural network and the HFS $\eta$ distributions in both simulations.
The model dependence of the unfolding for the HFS $\eta$ distribution is significantly larger than for the neural network distribution.
The size of the deviation from unity in the discrimination test for the neural network is large compared to the size of the unfolding model dependence.

%

\FloatBarrier
 \bibliographystyle{elsarticle-num} 
 \bibliography{cas-refs,HEPML,other-refs}

\begin{thebibliography}{10}
\expandafter\ifx\csname url\endcsname\relax
  \def\url#1{\texttt{#1}}\fi
\expandafter\ifx\csname urlprefix\endcsname\relax\def\urlprefix{URL }\fi
\expandafter\ifx\csname href\endcsname\relax
  \def\href#1#2{#2} \def\path#1{#1}\fi

\bibitem{CMS:2022kqg}
{ CMS} Collaboration, A.~Tumasyan, et~al., {Measurement of the differential
  $\hbox {t}\overline{\hbox {t}}$ production cross section as a function of the
  jet mass and extraction of the top quark mass in hadronic decays of boosted
  top quarks}, Eur. Phys. J. C 83~(7) (2023) 560.
\newblock \href {http://arxiv.org/abs/2211.01456} {\path{arXiv:2211.01456}},
  \href {https://doi.org/10.1140/epjc/s10052-023-11587-8}
  {\path{doi:10.1140/epjc/s10052-023-11587-8}}.

\bibitem{H1:2017bml}
{ H1} Collaboration, V.~Andreev, et~al., {Determination of the strong coupling
  constant $\alpha_s(m_Z)$ in next-to-next-to-leading order QCD using H1 jet
  cross section measurements}, Eur. Phys. J. C 77~(11) (2017) 791, [Erratum:
  Eur.Phys.J.C 81, 738 (2021)].
\newblock \href {http://arxiv.org/abs/1709.07251} {\path{arXiv:1709.07251}},
  \href {https://doi.org/10.1140/epjc/s10052-017-5314-7}
  {\path{doi:10.1140/epjc/s10052-017-5314-7}}.

\bibitem{Collaboration:2023xha}
{ ZEUS} Collaboration, Z.~Collaboration, {Measurement of jet production in deep
  inelastic scattering and NNLO determination of the strong coupling at ZEUS}
  (9 2023).
\newblock \href {http://arxiv.org/abs/2309.02889} {\path{arXiv:2309.02889}}.

\bibitem{H1:2015ubc}
{ H1, ZEUS} Collaboration, H.~Abramowicz, et~al., {Combination of measurements
  of inclusive deep inelastic ${e^{\pm }p}$ scattering cross sections and QCD
  analysis of HERA data}, Eur. Phys. J. C 75 (2015).
\newblock \href {http://arxiv.org/abs/1506.06042} {\path{arXiv:1506.06042}},
  \href {https://doi.org/10.1140/epjc/s10052-015-3710-4}
  {\path{doi:10.1140/epjc/s10052-015-3710-4}}.

\bibitem{Harland-Lang:2016yfn}
L.~A. Harland-Lang, A.~D. Martin, P.~Motylinski, R.~S. Thorne, {The impact of
  the final HERA combined data on PDFs obtained from a global fit}, Eur. Phys.
  J. C 76~(4) (2016) 186.
\newblock \href {http://arxiv.org/abs/1601.03413} {\path{arXiv:1601.03413}},
  \href {https://doi.org/10.1140/epjc/s10052-016-4020-1}
  {\path{doi:10.1140/epjc/s10052-016-4020-1}}.

\bibitem{Hou:2019efy}
T.-J. Hou, et~al., {New CTEQ global analysis of quantum chromodynamics with
  high-precision data from the LHC}, Phys. Rev. D 103~(1) (2021) 014013.
\newblock \href {http://arxiv.org/abs/1912.10053} {\path{arXiv:1912.10053}},
  \href {https://doi.org/10.1103/PhysRevD.103.014013}
  {\path{doi:10.1103/PhysRevD.103.014013}}.

\bibitem{NNPDF:2021njg}
{ NNPDF} Collaboration, R.~D. Ball, et~al., {The path to proton structure at
  1\% accuracy}, Eur. Phys. J. C 82~(5) (2022) 428.
\newblock \href {http://arxiv.org/abs/2109.02653} {\path{arXiv:2109.02653}},
  \href {https://doi.org/10.1140/epjc/s10052-022-10328-7}
  {\path{doi:10.1140/epjc/s10052-022-10328-7}}.

\bibitem{Campbell:2022qmc}
J.~M. Campbell, et~al., {Event Generators for High-Energy Physics Experiments},
  in: {Snowmass 2021}, 2022.
\newblock \href {http://arxiv.org/abs/2203.11110} {\path{arXiv:2203.11110}}.

\bibitem{Arratia:2022wny}
M.~Arratia, D.~Britzger, O.~Long, B.~Nachman, {Optimizing observables with
  machine learning for better unfolding}, JINST 17~(07) (2022) P07009.
\newblock \href {http://arxiv.org/abs/2203.16722} {\path{arXiv:2203.16722}},
  \href {https://doi.org/10.1088/1748-0221/17/07/P07009}
  {\path{doi:10.1088/1748-0221/17/07/P07009}}.

\bibitem{Datta:2018mwd}
K.~Datta, D.~Kar, D.~Roy, {Unfolding with Generative Adversarial Networks}
  (2018).
\newblock \href {http://arxiv.org/abs/1806.00433} {\path{arXiv:1806.00433}}.

\bibitem{bunse2018unification}
M.~Bunse, N.~Piatkowski, T.~Ruhe, W.~Rhode, K.~Morik, Unification of
  deconvolution algorithms for {C}herenkov astronomy, in: 5th International
  Conference on Data Science and Advanced Analytics (DSAA), IEEE, 2018, pp.
  21--30.

\bibitem{Ruhe2019MiningFS}
T.~Ruhe, T.~Voigt, M.~Wornowizki, M.~B{\"o}rner, W.~Rhode, K.~Morik, Mining for
  spectra - the dortmund spectrum estimation algorithm, 2019.

\bibitem{Andreassen:2019cjw}
A.~Andreassen, P.~T. Komiske, E.~M. Metodiev, B.~Nachman, J.~Thaler, {OmniFold:
  A Method to Simultaneously Unfold All Observables}, Phys. Rev. Lett. 124
  (2020) 182001.
\newblock \href {http://arxiv.org/abs/1911.09107} {\path{arXiv:1911.09107}},
  \href {https://doi.org/10.1103/PhysRevLett.124.182001}
  {\path{doi:10.1103/PhysRevLett.124.182001}}.

\bibitem{Bellagente:2019uyp}
M.~Bellagente, A.~Butter, G.~Kasieczka, T.~Plehn, R.~Winterhalder, {How to GAN
  away Detector Effects} (2019).
\newblock \href {http://arxiv.org/abs/1912.00477} {\path{arXiv:1912.00477}},
  \href {https://doi.org/10.21468/SciPostPhys.8.4.070}
  {\path{doi:10.21468/SciPostPhys.8.4.070}}.

\bibitem{1800956}
M.~Bellagente, A.~Butter, G.~Kasieczka, T.~Plehn, A.~Rousselot,
  R.~Winterhalder, {Invertible Networks or Partons to Detector and Back Again}
  (Jun 2020).
\newblock \href {http://arxiv.org/abs/2006.06685} {\path{arXiv:2006.06685}},
  \href {https://doi.org/10.21468/SciPostPhys.9.5.074}
  {\path{doi:10.21468/SciPostPhys.9.5.074}}.

\bibitem{Vandegar:2020yvw}
M.~Vandegar, M.~Kagan, A.~Wehenkel, G.~Louppe,
  \href{https://proceedings.mlr.press/v130/vandegar21a.html}{{Neural Empirical
  Bayes: Source Distribution Estimation and its Applications to
  Simulation-Based Inference}}, in: A.~Banerjee, K.~Fukumizu (Eds.),
  {Proceedings of The 24th International Conference on Artificial Intelligence
  and Statistics}, Vol. 130 of Proceedings of Machine Learning Research, PMLR,
  2021, pp. 2107--2115.
\newblock \href {http://arxiv.org/abs/2011.05836} {\path{arXiv:2011.05836}}.
\newline\urlprefix\url{https://proceedings.mlr.press/v130/vandegar21a.html}

\bibitem{Andreassen:2021zzk}
A.~Andreassen, P.~T. Komiske, E.~M. Metodiev, B.~Nachman, A.~Suresh, J.~Thaler,
  {Scaffolding Simulations with Deep Learning for High-dimensional
  Deconvolution}, in: {9th International Conference on Learning
  Representations}, 2021.
\newblock \href {http://arxiv.org/abs/2105.04448} {\path{arXiv:2105.04448}}.

\bibitem{Howard:2021pos}
J.~N. Howard, S.~Mandt, D.~Whiteson, Y.~Yang, {Foundations of a Fast,
  Data-Driven, Machine-Learned Simulator} (Jan 2021).
\newblock \href {http://arxiv.org/abs/2101.08944} {\path{arXiv:2101.08944}}.

\bibitem{Backes:2022vmn}
M.~Backes, A.~Butter, M.~Dunford, B.~Malaescu, {An unfolding method based on
  conditional Invertible Neural Networks (cINN) using iterative training} (12
  2022).
\newblock \href {http://arxiv.org/abs/2212.08674} {\path{arXiv:2212.08674}}.

\bibitem{Chan:2023tbf}
J.~Chan, B.~Nachman, {Unbinned profiled unfolding}, Phys. Rev. D 108~(1) (2023)
  016002.
\newblock \href {http://arxiv.org/abs/2302.05390} {\path{arXiv:2302.05390}},
  \href {https://doi.org/10.1103/PhysRevD.108.016002}
  {\path{doi:10.1103/PhysRevD.108.016002}}.

\bibitem{Shmakov:2023kjj}
A.~Shmakov, K.~Greif, M.~Fenton, A.~Ghosh, P.~Baldi, D.~Whiteson, {End-To-End
  Latent Variational Diffusion Models for Inverse Problems in High Energy
  Physics} (5 2023).
\newblock \href {http://arxiv.org/abs/2305.10399} {\path{arXiv:2305.10399}}.

\bibitem{Alghamdi:2023emm}
T.~Alghamdi, et~al., {Toward a generative modeling analysis of CLAS exclusive
  $2\pi$ photoproduction} (7 2023).
\newblock \href {http://arxiv.org/abs/2307.04450} {\path{arXiv:2307.04450}}.

\bibitem{H1:2021wkz}
{ H1} Collaboration, V.~Andreev, et~al., {Measurement of lepton-jet correlation
  in deep-inelastic scattering with the H1 detector using machine learning for
  unfolding} (Aug 2021).
\newblock \href {http://arxiv.org/abs/2108.12376} {\path{arXiv:2108.12376}}.

\bibitem{H1prelim-22-031}
{H1 Collaboration},
  \href{https://www-h1.desy.de/h1/www/publications/htmlsplit/H1prelim-22-031.long.html}{{Machine
  learning-assisted measurement of multi-differential lepton-jet correlations
  in deep-inelastic scattering with the H1 detector}}, H1prelim-22-031 (2022).
\newline\urlprefix\url{https://www-h1.desy.de/h1/www/publications/htmlsplit/H1prelim-22-031.long.html}

\bibitem{H1:2023fzk}
{ H1} Collaboration, V.~Andreev, et~al., {Unbinned Deep Learning Jet
  Substructure Measurement in High $Q^2$ ep collisions at HERA} (3 2023).
\newblock \href {http://arxiv.org/abs/2303.13620} {\path{arXiv:2303.13620}}.

\bibitem{H1prelim-21-031}
{H1 Collaboration},
  \href{https://www-h1.desy.de/h1/www/publications/htmlsplit/H1prelim-21-031.long.html}{{Measurement
  of lepton-jet correlations in high $Q^2$ neutral-current DIS with the H1
  detector at HERA}}, H1prelim-21-031 (2021).
\newline\urlprefix\url{https://www-h1.desy.de/h1/www/publications/htmlsplit/H1prelim-21-031.long.html}

\bibitem{LHCb:2022rky}
{ LHCb} Collaboration, {Multidifferential study of identified charged hadron
  distributions in $Z$-tagged jets in proton-proton collisions at $\sqrt{s}=$13
  TeV} (8 2022).
\newblock \href {http://arxiv.org/abs/2208.11691} {\path{arXiv:2208.11691}}.

\bibitem{Komiske:2022vxg}
P.~T. Komiske, S.~Kryhin, J.~Thaler, {Disentangling quarks and gluons in CMS
  open data}, Phys. Rev. D 106~(9) (2022) 094021.
\newblock \href {http://arxiv.org/abs/2205.04459} {\path{arXiv:2205.04459}},
  \href {https://doi.org/10.1103/PhysRevD.106.094021}
  {\path{doi:10.1103/PhysRevD.106.094021}}.

\bibitem{Song:2023sxb}
{ STAR} Collaboration, Y.~Song, {Measurement of CollinearDrop jet mass and its
  correlation with SoftDrop groomed jet substructure observables in
  $\sqrt{s}=200$ GeV $pp$ collisions by STAR} (7 2023).
\newblock \href {http://arxiv.org/abs/2307.07718} {\path{arXiv:2307.07718}}.

\bibitem{Arratia:2021otl}
M.~Arratia, et~al., {Publishing unbinned differential cross section results},
  JINST 17~(01) (2022) P01024.
\newblock \href {http://arxiv.org/abs/2109.13243} {\path{arXiv:2109.13243}},
  \href {https://doi.org/10.1088/1748-0221/17/01/P01024}
  {\path{doi:10.1088/1748-0221/17/01/P01024}}.

\bibitem{Brehmer:2018kdj}
J.~Brehmer, K.~Cranmer, G.~Louppe, J.~Pavez, {Constraining Effective Field
  Theories with Machine Learning}, Phys. Rev. Lett. 121~(11) (2018) 111801.
\newblock \href {http://arxiv.org/abs/1805.00013} {\path{arXiv:1805.00013}},
  \href {https://doi.org/10.1103/PhysRevLett.121.111801}
  {\path{doi:10.1103/PhysRevLett.121.111801}}.

\bibitem{Brehmer:2018eca}
J.~Brehmer, K.~Cranmer, G.~Louppe, J.~Pavez, {A Guide to Constraining Effective
  Field Theories with Machine Learning}, Phys. Rev. D 98~(5) (2018) 052004.
\newblock \href {http://arxiv.org/abs/1805.00020} {\path{arXiv:1805.00020}},
  \href {https://doi.org/10.1103/PhysRevD.98.052004}
  {\path{doi:10.1103/PhysRevD.98.052004}}.

\bibitem{Brehmer:2018hga}
J.~Brehmer, G.~Louppe, J.~Pavez, K.~Cranmer, {Mining gold from implicit models
  to improve likelihood-free inference}, Proc. Nat. Acad. Sci. 117~(10) (2020)
  5242--5249.
\newblock \href {http://arxiv.org/abs/1805.12244} {\path{arXiv:1805.12244}},
  \href {https://doi.org/10.1073/pnas.1915980117}
  {\path{doi:10.1073/pnas.1915980117}}.

\bibitem{DeCastro:2018psv}
P.~De~Castro, T.~Dorigo, {INFERNO: Inference-Aware Neural Optimisation},
  Comput. Phys. Commun. 244 (2019) 170--179.
\newblock \href {http://arxiv.org/abs/1806.04743} {\path{arXiv:1806.04743}},
  \href {https://doi.org/10.1016/j.cpc.2019.06.007}
  {\path{doi:10.1016/j.cpc.2019.06.007}}.

\bibitem{Elwood:2018qsr}
A.~Elwood, D.~Kr\"ucker, {Direct optimisation of the discovery significance
  when training neural networks to search for new physics in particle
  colliders} (6 2018).
\newblock \href {http://arxiv.org/abs/1806.00322} {\path{arXiv:1806.00322}}.

\bibitem{Andreassen:2019nnm}
A.~Andreassen, B.~Nachman, {Neural Networks for Full Phase-space Reweighting
  and Parameter Tuning}, Phys. Rev. D 101~(9) (2020) 091901.
\newblock \href {http://arxiv.org/abs/1907.08209} {\path{arXiv:1907.08209}},
  \href {https://doi.org/10.1103/PhysRevD.101.091901}
  {\path{doi:10.1103/PhysRevD.101.091901}}.

\bibitem{Brehmer:2019xox}
J.~Brehmer, F.~Kling, I.~Espejo, K.~Cranmer, {MadMiner: Machine learning-based
  inference for particle physics}, Comput. Softw. Big Sci. 4~(1) (2020) 3.
\newblock \href {http://arxiv.org/abs/1907.10621} {\path{arXiv:1907.10621}},
  \href {https://doi.org/10.1007/s41781-020-0035-2}
  {\path{doi:10.1007/s41781-020-0035-2}}.

\bibitem{Wunsch:2020iuh}
S.~Wunsch, S.~J\"orger, R.~Wolf, G.~Quast, {Optimal Statistical Inference in
  the Presence of Systematic Uncertainties Using Neural Network Optimization
  Based on Binned Poisson Likelihoods with Nuisance Parameters}, Comput. Softw.
  Big Sci. 5~(1) (2021) 4.
\newblock \href {http://arxiv.org/abs/2003.07186} {\path{arXiv:2003.07186}},
  \href {https://doi.org/10.1007/s41781-020-00049-5}
  {\path{doi:10.1007/s41781-020-00049-5}}.

\bibitem{Ghosh:2021roe}
A.~Ghosh, B.~Nachman, D.~Whiteson, {Uncertainty-aware machine learning for high
  energy physics}, Phys. Rev. D 104~(5) (2021) 056026.
\newblock \href {http://arxiv.org/abs/2105.08742} {\path{arXiv:2105.08742}},
  \href {https://doi.org/10.1103/PhysRevD.104.056026}
  {\path{doi:10.1103/PhysRevD.104.056026}}.

\bibitem{GomezAmbrosio:2022mpm}
R.~Gomez~Ambrosio, J.~ter Hoeve, M.~Madigan, J.~Rojo, V.~Sanz, {Unbinned
  multivariate observables for global SMEFT analyses from machine learning},
  JHEP 03 (2023) 033.
\newblock \href {http://arxiv.org/abs/2211.02058} {\path{arXiv:2211.02058}},
  \href {https://doi.org/10.1007/JHEP03(2023)033}
  {\path{doi:10.1007/JHEP03(2023)033}}.

\bibitem{Simpson:2022suz}
N.~Simpson, L.~Heinrich, {neos: End-to-End-Optimised Summary Statistics for
  High Energy Physics}, J. Phys. Conf. Ser. 2438~(1) (2023) 012105.
\newblock \href {http://arxiv.org/abs/2203.05570} {\path{arXiv:2203.05570}},
  \href {https://doi.org/10.1088/1742-6596/2438/1/012105}
  {\path{doi:10.1088/1742-6596/2438/1/012105}}.

\bibitem{Blance:2019ibf}
A.~Blance, M.~Spannowsky, P.~Waite, {Adversarially-trained autoencoders for
  robust unsupervised new physics searches}, JHEP 10 (2019) 047.
\newblock \href {http://arxiv.org/abs/1905.10384} {\path{arXiv:1905.10384}},
  \href {https://doi.org/10.1007/JHEP10(2019)047}
  {\path{doi:10.1007/JHEP10(2019)047}}.

\bibitem{Englert:2018cfo}
C.~Englert, P.~Galler, P.~Harris, M.~Spannowsky, {Machine Learning
  Uncertainties with Adversarial Neural Networks}, Eur. Phys. J. C79~(1) (2019)
  4.
\newblock \href {http://arxiv.org/abs/1807.08763} {\path{arXiv:1807.08763}},
  \href {https://doi.org/10.1140/epjc/s10052-018-6511-8}
  {\path{doi:10.1140/epjc/s10052-018-6511-8}}.

\bibitem{Louppe:2016ylz}
G.~Louppe, M.~Kagan, K.~Cranmer,
  \href{https://papers.nips.cc/paper/2017/hash/48ab2f9b45957ab574cf005eb8a76760-Abstract.html}{{Learning
  to Pivot with Adversarial Networks}}, in: I.~Guyon, U.~V. Luxburg, S.~Bengio,
  H.~Wallach, R.~Fergus, S.~Vishwanathan, R.~Garnett (Eds.), {Advances in
  Neural Information Processing Systems}, Vol.~30, Curran Associates, Inc.,
  2017.
\newblock \href {http://arxiv.org/abs/1611.01046} {\path{arXiv:1611.01046}}.
\newline\urlprefix\url{https://papers.nips.cc/paper/2017/hash/48ab2f9b45957ab574cf005eb8a76760-Abstract.html}

\bibitem{Dolen:2016kst}
J.~Dolen, P.~Harris, S.~Marzani, S.~Rappoccio, N.~Tran, {Thinking outside the
  ROCs: Designing Decorrelated Taggers (DDT) for jet substructure}, JHEP 05
  (2016) 156.
\newblock \href {http://arxiv.org/abs/1603.00027} {\path{arXiv:1603.00027}},
  \href {https://doi.org/10.1007/JHEP05(2016)156}
  {\path{doi:10.1007/JHEP05(2016)156}}.

\bibitem{Moult:2017okx}
I.~Moult, B.~Nachman, D.~Neill, {Convolved Substructure: Analytically
  Decorrelating Jet Substructure Observables} (2017).
\newblock \href {http://arxiv.org/abs/1710.06859} {\path{arXiv:1710.06859}}.

\bibitem{Stevens:2013dya}
J.~Stevens, M.~Williams, {uBoost: A boosting method for producing uniform
  selection efficiencies from multivariate classifiers}, JINST 8 (2013) P12013.
\newblock \href {http://arxiv.org/abs/1305.7248} {\path{arXiv:1305.7248}},
  \href {https://doi.org/10.1088/1748-0221/8/12/P12013}
  {\path{doi:10.1088/1748-0221/8/12/P12013}}.

\bibitem{Shimmin:2017mfk}
C.~Shimmin, P.~Sadowski, P.~Baldi, E.~Weik, D.~Whiteson, E.~Goul, A.~Sogaard,
  {Decorrelated Jet Substructure Tagging using Adversarial Neural Networks},
  Phys. Rev. D96~(7) (2017) 074034.
\newblock \href {http://arxiv.org/abs/1703.03507} {\path{arXiv:1703.03507}},
  \href {https://doi.org/10.1103/PhysRevD.96.074034}
  {\path{doi:10.1103/PhysRevD.96.074034}}.

\bibitem{Bradshaw:2019ipy}
L.~Bradshaw, R.~K. Mishra, A.~Mitridate, B.~Ostdiek, {Mass Agnostic Jet
  Taggers} (2019).
\newblock \href {http://arxiv.org/abs/1908.08959} {\path{arXiv:1908.08959}},
  \href {https://doi.org/10.21468/SciPostPhys.8.1.011}
  {\path{doi:10.21468/SciPostPhys.8.1.011}}.

\bibitem{ATL-PHYS-PUB-2018-014}
\href{https://cds.cern.ch/record/2630973}{{Performance of mass-decorrelated jet
  substructure observables for hadronic two-body decay tagging in ATLAS}},
  Tech. Rep. ATL-PHYS-PUB-2018-014, CERN, Geneva (Jul 2018).
\newline\urlprefix\url{https://cds.cern.ch/record/2630973}

\bibitem{DiscoFever}
G.~Kasieczka, D.~Shih, {DisCo Fever: Robust Networks Through Distance
  Correlation} (2020).
\newblock \href {http://arxiv.org/abs/2001.05310} {\path{arXiv:2001.05310}},
  \href {https://doi.org/10.1103/PhysRevLett.125.122001}
  {\path{doi:10.1103/PhysRevLett.125.122001}}.

\bibitem{Wunsch:2019qbo}
S.~Wunsch, S.~J\'{o}rger, R.~Wolf, G.~Quast, {Reducing the dependence of the
  neural network function to systematic uncertainties in the input space}
  (2019).
\newblock \href {http://arxiv.org/abs/1907.11674} {\path{arXiv:1907.11674}},
  \href {https://doi.org/10.1007/s41781-020-00037-9}
  {\path{doi:10.1007/s41781-020-00037-9}}.

\bibitem{Rogozhnikov:2014zea}
A.~Rogozhnikov, A.~Bukva, V.~Gligorov, A.~Ustyuzhanin, M.~Williams, {New
  approaches for boosting to uniformity}, JINST 10~(03) (2015) T03002.
\newblock \href {http://arxiv.org/abs/1410.4140} {\path{arXiv:1410.4140}},
  \href {https://doi.org/10.1088/1748-0221/10/03/T03002}
  {\path{doi:10.1088/1748-0221/10/03/T03002}}.

\bibitem{10.1088/2632-2153/ab9023}
C.~Collaboration, {A deep neural network to search for new long-lived particles
  decaying to jets}, Machine Learning: Science and Technology (2020).
\newblock \href {http://arxiv.org/abs/1912.12238} {\path{arXiv:1912.12238}},
  \href {https://doi.org/10.1088/2632-2153/ab9023}
  {\path{doi:10.1088/2632-2153/ab9023}}.

\bibitem{Kasieczka:2020pil}
G.~Kasieczka, B.~Nachman, M.~D. Schwartz, D.~Shih, {ABCDisCo: Automating the
  ABCD Method with Machine Learning} (Jul 2020).
\newblock \href {http://arxiv.org/abs/2007.14400} {\path{arXiv:2007.14400}},
  \href {https://doi.org/10.1103/PhysRevD.103.035021}
  {\path{doi:10.1103/PhysRevD.103.035021}}.

\bibitem{Kitouni:2020xgb}
O.~Kitouni, B.~Nachman, C.~Weisser, M.~Williams, {Enhancing searches for
  resonances with machine learning and moment decomposition} (Oct 2020).
\newblock \href {http://arxiv.org/abs/2010.09745} {\path{arXiv:2010.09745}}.

\bibitem{Estrade:2019gzk}
V.~Estrade, C.~Germain, I.~Guyon, D.~Rousseau, {Systematic aware learning - A
  case study in High Energy Physics}, EPJ Web Conf. 214 (2019) 06024.
\newblock \href {https://doi.org/10.1051/epjconf/201921406024}
  {\path{doi:10.1051/epjconf/201921406024}}.

\bibitem{Aguilar-Saavedra:2017rzt}
J.~A. Aguilar-Saavedra, J.~H. Collins, R.~K. Mishra, {A generic anti-QCD jet
  tagger}, JHEP 11 (2017) 163.
\newblock \href {http://arxiv.org/abs/1709.01087} {\path{arXiv:1709.01087}},
  \href {https://doi.org/10.1007/JHEP11(2017)163}
  {\path{doi:10.1007/JHEP11(2017)163}}.

\bibitem{aguilarsaavedra2020mass}
J.~A. Aguilar-Saavedra, F.~R. Joaquim, J.~F. Seabra, {Mass Unspecific
  Supervised Tagging (MUST) for boosted jets} (2020).
\newblock \href {http://arxiv.org/abs/2008.12792} {\path{arXiv:2008.12792}},
  \href {https://doi.org/10.1007/JHEP03(2021)012}
  {\path{doi:10.1007/JHEP03(2021)012}}.

\bibitem{Aguilar-Saavedra:2023pde}
J.~A. Aguilar-Saavedra, E.~Arganda, F.~R. Joaquim, R.~M. Sand\'a~Seoane, J.~F.
  Seabra, {Gradient Boosting MUST taggers for highly-boosted jets} (5 2023).
\newblock \href {http://arxiv.org/abs/2305.04957} {\path{arXiv:2305.04957}}.

\bibitem{chollet2015keras}
F.~Chollet, Keras, \url{https://github.com/fchollet/keras} (2015).

\bibitem{tensorflow}
M.~Abadi, P.~Barham, J.~Chen, Z.~Chen, A.~Davis, J.~Dean, M.~Devin,
  S.~Ghemawat, G.~Irving, M.~Isard, et~al., Tensorflow: A system for
  large-scale machine learning., in: OSDI, Vol.~16, 2016, pp. 265--283.

\bibitem{adam}
D.~Kingma, J.~Ba, Adam: A method for stochastic optimization (2014).
\newblock \href {http://arxiv.org/abs/1412.6980} {\path{arXiv:1412.6980}}.

\bibitem{Schmitt:2012kp}
S.~Schmitt, {TUnfold: an algorithm for correcting migration effects in high
  energy physics}, JINST 7 (2012) T10003.
\newblock \href {http://arxiv.org/abs/1205.6201} {\path{arXiv:1205.6201}},
  \href {https://doi.org/10.1088/1748-0221/7/10/T10003}
  {\path{doi:10.1088/1748-0221/7/10/T10003}}.

\bibitem{Brun:1997pa}
R.~Brun, F.~Rademakers, {ROOT: An object oriented data analysis framework},
  Nucl. Instrum. Meth. A 389 (1997) 81--86.
\newblock \href {https://doi.org/10.1016/S0168-9002(97)00048-X}
  {\path{doi:10.1016/S0168-9002(97)00048-X}}.

\bibitem{H1:1996jzy}
{ H1} Collaboration, I.~Abt, et~al., {The Tracking, calorimeter and muon
  detectors of the H1 experiment at HERA}, Nucl. Instrum. Meth. A 386 (1997)
  348--396.
\newblock \href {https://doi.org/10.1016/S0168-9002(96)00894-7}
  {\path{doi:10.1016/S0168-9002(96)00894-7}}.

\bibitem{H1:1996prr}
{ H1} Collaboration, I.~Abt, et~al., {The H1 detector at HERA}, Nucl. Instrum.
  Meth. A 386 (1997) 310--347.
\newblock \href {https://doi.org/10.1016/S0168-9002(96)00893-5}
  {\path{doi:10.1016/S0168-9002(96)00893-5}}.

\bibitem{Arratia:2021tsq}
M.~Arratia, D.~Britzger, O.~Long, B.~Nachman, {Reconstructing the kinematics of
  deep inelastic scattering with deep learning}, Nucl. Instrum. Meth. A 1025
  (2022) 166164.
\newblock \href {http://arxiv.org/abs/2110.05505} {\path{arXiv:2110.05505}},
  \href {https://doi.org/10.1016/j.nima.2021.166164}
  {\path{doi:10.1016/j.nima.2021.166164}}.

\bibitem{Jung:1993gf}
H.~Jung, {Hard diffractive scattering in high-energy e p collisions and the
  Monte Carlo generator RAPGAP}, Comput. Phys. Commun. 86 (1995) 147--161.
\newblock \href {https://doi.org/10.1016/0010-4655(94)00150-Z}
  {\path{doi:10.1016/0010-4655(94)00150-Z}}.

\bibitem{Charchula:1994kf}
K.~Charchula, G.~A. Schuler, H.~Spiesberger, {Combined QED and QCD radiative
  effects in deep inelastic lepton - proton scattering: The Monte Carlo
  generator DJANGO6}, Comput. Phys. Commun. 81 (1994) 381--402.
\newblock \href {https://doi.org/10.1016/0010-4655(94)90086-8}
  {\path{doi:10.1016/0010-4655(94)90086-8}}.

\bibitem{Spiesberger:237380}
H.~Spiesberger, et~al., \href{https://cds.cern.ch/record/237380}{{Radiative
  corrections at HERA}} (1992) 798--839.
\newline\urlprefix\url{https://cds.cern.ch/record/237380}

\bibitem{Kwiatkowski:1990cx}
A.~Kwiatkowski, H.~Spiesberger, H.~J. Mohring, {Characteristics of radiative
  events in deep inelastic \emph{ep} scattering at HERA}, Z. Phys. C 50 (1991)
  165--178.
\newblock \href {https://doi.org/10.1007/BF01558572}
  {\path{doi:10.1007/BF01558572}}.

\bibitem{Kwiatkowski:1990es}
A.~Kwiatkowski, H.~Spiesberger, H.~J. Mohring, {Heracles: An Event Generator
  for $e p$ Interactions at {HERA} Energies Including Radiative Processes:
  Version 1.0}, Comput. Phys. Commun. 69 (1992) 155--172.
\newblock \href {https://doi.org/10.1016/0010-4655(92)90136-M}
  {\path{doi:10.1016/0010-4655(92)90136-M}}.

\bibitem{geant4}
S.~Agostinelli, et~al.,
  \href{http://www.sciencedirect.com/science/article/pii/S0168900203013688}{Geant4
  - a simulation toolkit}, Nuclear Instruments and Methods in Physics Research
  Section A: Accelerators, Spectrometers, Detectors and Associated Equipment
  506~(3) (2003) 250 -- 303.
\newblock \href {https://doi.org/https://doi.org/10.1016/S0168-9002(03)01368-8}
  {\path{doi:https://doi.org/10.1016/S0168-9002(03)01368-8}}.
\newline\urlprefix\url{http://www.sciencedirect.com/science/article/pii/S0168900203013688}

\bibitem{energyflowthesis}
M.~Peez, {Search for deviations from the standard model in high transverse
  energy processes at the electron proton collider HERA. (Thesis, Univ. Lyon)},
  Ph.D. thesis (Jun 2003).

\bibitem{energyflowthesis2}
S.~Hellwig, {Untersuchung der $D^*$ - $\pi$ slow Double Tagging Methode in
  Charmanalysen}, {Diploma thesis, Univ. Hamburg} (Jun 2004).

\bibitem{energyflowthesis3}
B.~Portheault, {First measurement of charged and neutral current cross sections
  with the polarized positron beam at HERA II and QCD-electroweak analyses.
  (Thesis, Univ. Paris XI)}, Ph.D. thesis (Mar 2005).

\bibitem{H1:2012qti}
{ H1} Collaboration, F.~D. Aaron, et~al., {Inclusive Deep Inelastic Scattering
  at High $Q^2$ with Longitudinally Polarised Lepton Beams at HERA}, JHEP 09
  (2012) 061.
\newblock \href {http://arxiv.org/abs/1206.7007} {\path{arXiv:1206.7007}},
  \href {https://doi.org/10.1007/JHEP09(2012)061}
  {\path{doi:10.1007/JHEP09(2012)061}}.

\bibitem{H1:2014cbm}
{ H1} Collaboration, V.~Andreev, et~al., {Measurement of multijet production in
  $ep$ collisions at high $Q^2$ and determination of the strong coupling
  $\alpha _s$}, Eur. Phys. J. C 75~(2) (2015) 65.
\newblock \href {http://arxiv.org/abs/1406.4709} {\path{arXiv:1406.4709}},
  \href {https://doi.org/10.1140/epjc/s10052-014-3223-6}
  {\path{doi:10.1140/epjc/s10052-014-3223-6}}.

\bibitem{Britzger:2021xcx}
D.~Britzger, S.~Levonian, S.~Schmitt, D.~South, {Preservation through
  modernisation: The software of the H1 experiment at HERA}, EPJ Web Conf. 251
  (2021) 03004.
\newblock \href {http://arxiv.org/abs/2106.11058} {\path{arXiv:2106.11058}},
  \href {https://doi.org/10.1051/epjconf/202125103004}
  {\path{doi:10.1051/epjconf/202125103004}}.

\end{thebibliography}

\end{document}